\documentclass{aa}
\renewcommand{\arraystretch}{1.1}
\usepackage{color}
\usepackage{array}
\usepackage{url}
\usepackage[bottom]{footmisc}
\usepackage{siunitx}
\usepackage{xspace}

\newcommand{\nustar}{\textit{NuSTAR}\xspace}
\newcommand{\xmm}{\textit{XMM-Newton}\xspace}
\newcommand{\slxtres}{{SLX~1744$-$300}\xspace}
\newcommand{\slxdos}{{SLX~1744$-$299}\xspace}

\newcommand{\erg}{\mathrm{erg}}
\newcommand{\cm}{\mathrm{cm}}
\newcommand{\s}{\mathrm{s}}
\newcommand{\m}{\mathrm{m}}

\newcommand{\fb}{\mathrm{erg~cm}^{-2}}
\newcommand{\lum}{\mathrm{erg~s}^{-1}}
\newcommand{\flux}{\mathrm{erg~cm}^{-2}~\mathrm{s}^{-1}}

\usepackage{graphicx}
\usepackage{multirow}
\usepackage{booktabs}
\usepackage{caption}
\usepackage{subcaption}
\usepackage{adjustbox}
\usepackage[toc,page]{appendix}
\usepackage{amsmath}
\usepackage{nccmath}
\usepackage[flushleft]{threeparttable}

\usepackage{txfonts}
\usepackage[]{hyperref}
\hypersetup{unicode=true, colorlinks=true, linkcolor=[rgb]{0.53, 0.15, 0.34}, citecolor=blue, filecolor=[rgb]{1.0, 0.13, 0.32}, urlcolor=[rgb]{0.53, 0.15, 0.34}}

\usepackage{ulem}

\begin{document} 



   \title{Resolving \slxdos\ and \slxtres\ in the hard X-ray band: Implications for their ultracompact nature}

 \subtitle{}

   \author{Enzo A. Saavedra\inst{1,2} 
           \and
           Montserrat Armas Padilla\inst{1,2}
           \and
           Teo Muñoz-Darias\inst{1,2} 
           }

   \institute{
   Instituto de Astrofísica de Canarias (IAC), Vía Láctea s/n, La Laguna 38205, S/C de Tenerife, Spain \and
   Departamento de Astrofísica, Universidad de La Laguna, La Laguna, E-38205, S/C de Tenerife, Spain
   }

   \date{Received; accepted}

 
   \abstract
{Persistent, low-luminosity low-mass X-ray binaries offer a unique opportunity to study accretion in this poorly understood regime as well as to unveil new members of the ultracompact X-ray binary (UCXB) family, which are characterised by orbital periods ($P_{\rm orb}$) shorter than $\sim 80$~min.  
We report on a \textit{NuSTAR} archival observation that, for the first time above 10~keV, spatially resolves the Galactic Centre pair SLX\,1744$-$299 and SLX\,1744$-$300.  We find \slxtres\ to be slightly brighter, with a flux ratio of $\sim 1.15$, increasing to $\sim 1.3$ when extrapolated to 0.5--10~keV.  Both the timing (root-mean-square variability) and spectral properties (well described in both cases by a thermal Comptonisation model) indicate that the systems were in the hard state.  The two sources, however, display markedly different behaviour throughout the observation. \slxdos\ shows a gradual flux decline consistent with a decrease in the mass-accretion rate, whereas \slxtres\ remains steady but exhibits two short-recurrence Type-I X-ray bursts indicative of mixed H and He burning.  

Combining our results with previously reported upper limits on the distance, we derived low persistent X-ray luminosities of $L_{\rm X}\lesssim 1.1\times10^{36}$~erg~s$^{-1}$ and $L_{\rm X}\lesssim 2.6\times10^{36}$~erg~s$^{-1}$ (3--78 keV) for \slxdos\ and \slxtres, respectively. The corresponding mass-accretion rates, when compared with the critical values from the disc instability model, favour $P_{\rm orb}\lesssim 90$~min and $P_{\rm orb}\lesssim 105$--155~min. Although both limits are formally compatible with the UCXB regime, the case of \slxdos\ appears significantly more compelling, also considering the previously reported intermediate-duration burst.}

   \titlerunning{A hard X-ray view of \slxdos and \slxtres}
   \authorrunning{Saavedra et al.}
   \maketitle
%
\section{Introduction}

Low-mass X-ray binaries (LMXBs) are stellar systems in which a compact object, either a black hole or a neutron star (NS), accretes matter from a low-mass companion star \citep[$\lesssim1{\rm M}_\odot$; see][for a recent review]{Bahramian2023hxga.book..120B}. To date, approximately 340 such systems have been identified in our Galaxy \citep[see e.g.,][]{Fortin2024}.

Of particular interest among LMXBs are ultracompact X-ray binaries (UCXBs), which are characterised by very short orbital periods ($P_{\rm orb} \lesssim 80$~min). In these tight systems, the companion is a degenerate or semi-degenerate star that no longer burns hydrogen in its core and may consist of electron-degenerate matter \citep[e.g.][]{Paczynski1981ApJ...248L..27P, Rappaport1982ApJ...254..616R, Verbunt1995xrbi.nasa..457V}. Therefore, these systems serve as unique laboratories for studying accretion processes in hydrogen-poor environments \citep[e.g.][]{Nelemans2010NewAR..54...87N}. Additionally, UCXBs are expected to be strong sources of low-frequency gravitational waves, making them key targets for the upcoming Laser Interferometer Space Antenna (LISA) mission \citep[e.g.][]{Nelemans2018arXiv180701060N, Tauris2018PhRvL.121m1105T, Chen2021MNRAS.503.3540C, Amaro-Seoane2023LRR....26....2A}.

Currently, there are only 20 confirmed UCXBs (i.e. with a measured orbital period shorter than 80 min) and 25 UCXB candidates \citep[see the online version of the UltraCompCAT catalogue\footnote{\url{https://research.iac.es/proyecto/compactos/UltraCompCAT}};][]{ArmasPadilla2023A&A...677A.186A}. Yet, the estimated number of UCXBs expected to exist in our Galaxy is $(0.2-1.9)\times10^5$ \citep{Zhu2012RAA....12.1526Z, VanHaaften2013A&A...552A..69V}. Measuring $P_{\rm orb}$ in these systems is observationally challenging. While X-ray timing techniques can be applied in systems hosting pulsars or showing eclipses or dips, most sources lack such features. In the optical, when a counterpart is detected, it is typically intrinsically faint, often requiring exposure times comparable to or longer than the orbital period. However, in the absence of a measured orbital period, indirect diagnostics can be used to identify UCXB candidates. These include signatures of the degenerate nature of the donor star, such as the presence or absence of specific emission and absorption features in the source spectra, as well as the properties of thermonuclear X-ray bursts (e.g. duration, recurrence time, and radiated energy), which provide insights into the composition of the accreted fuel. Likewise, the small size of the accretion disc in UCXBs results in lower optical-to-X-ray flux ratios than those associated with regular LMXBs, as the reprocessing region responsible for the optical emission is also reduced \citep[for more details, see][]{ArmasPadilla2023A&A...677A.186A}.
Last but not least, small discs can remain fully ionised when accreting at low rates, as predicted by the disc instability model \citep[DIM;][]{Lasota2001NewAR..45..449L}, allowing UCXBs to sustain persistent accretion even at X-ray luminosities as low as $L_{\mathrm{X}} \lesssim 10^{36}~\lum$ \citep[see, e.g.,][]{intZand2007A&A...465..953I}. This makes persistently accreting, low-luminosity LMXBs strong UCXB candidates.

\slxdos\ and \slxtres\ are two LMXBs located $\sim$2.7~arcmin apart and $\sim$2.3~arcmin from the pulsar PSR J1747$-$2958 (known as ‘the Mouse’; \citealt{Skinner1990MNRAS.243...72S, Pavlinsky2021A&A...650A..42P}). Owing to their small angular separation, they were initially detected as a single X-ray source with the telescope aboard Spacelab 2 (\citealt{Skinner1987Natur.330..544S}, see also \citealt{Kawai1988ApJ...330..130K}), and were only identified as two distinct objects a few years later through improved image reconstruction techniques \citep{Skinner1990MNRAS.243...72S}. Both LMXBs exhibit persistent fluxes of $\sim 10^{-10}~\flux$ in the 2--10~keV range, with the flux of \slxdos\ being approximately twice that of \slxtres\ \citep{Mori2005AdSpR..35.1137M}. 
However, long-term monitoring of the field with the \textit{Rossi} X-ray Timing Explorer showed that the unresolved emission varies by roughly a factor of two, hinting at the possibility that either or both sources are variable \citep{intZand2007A&A...465..953I}.
Type-I thermonuclear X-ray bursts have been observed in both systems, confirming the NS nature of their compact objects \citep{Skinner1990MNRAS.243...72S, Pavlinsky1994ApJ, Mori2005AdSpR..35.1137M, Galloway2008ApJS..179..360G}. In particular, \slxdos\ has exhibited several intermediate-duration bursts, which typically last a few minutes \citep{Pavlinsky1994ApJ, intZand2007A&A...465..953I, Alizai2020MNRAS.494.2509A}. None of the bursts from either system has shown photospheric radius expansion, so only upper limits on their distances have been estimated: $7.2 \pm 1.4$~kpc for \slxdos\ and $10.3 \pm 0.8$~kpc for \slxtres\ \citep{Chelovekov2017AstL}. These values translate into low persistent luminosities ($\lesssim 10^{36}~\lum$) in both cases. This, together with the presence of intermediate-duration bursts, has led to the classification of \slxdos\ as a UCXB candidate \citep{intZand2007A&A...465..953I}. The nature of \slxtres, on the other hand, remains largely unconstrained.

In this paper we present a detailed X-ray spectral and temporal analysis of a \nustar\ observation that, for the first time, resolves both systems individually in the hard X-ray band above 10 keV. 
This allowed us to characterise their hard X-ray spectra and further investigate their nature.

\begin{figure}
    \centering
    \includegraphics[width=1\columnwidth]{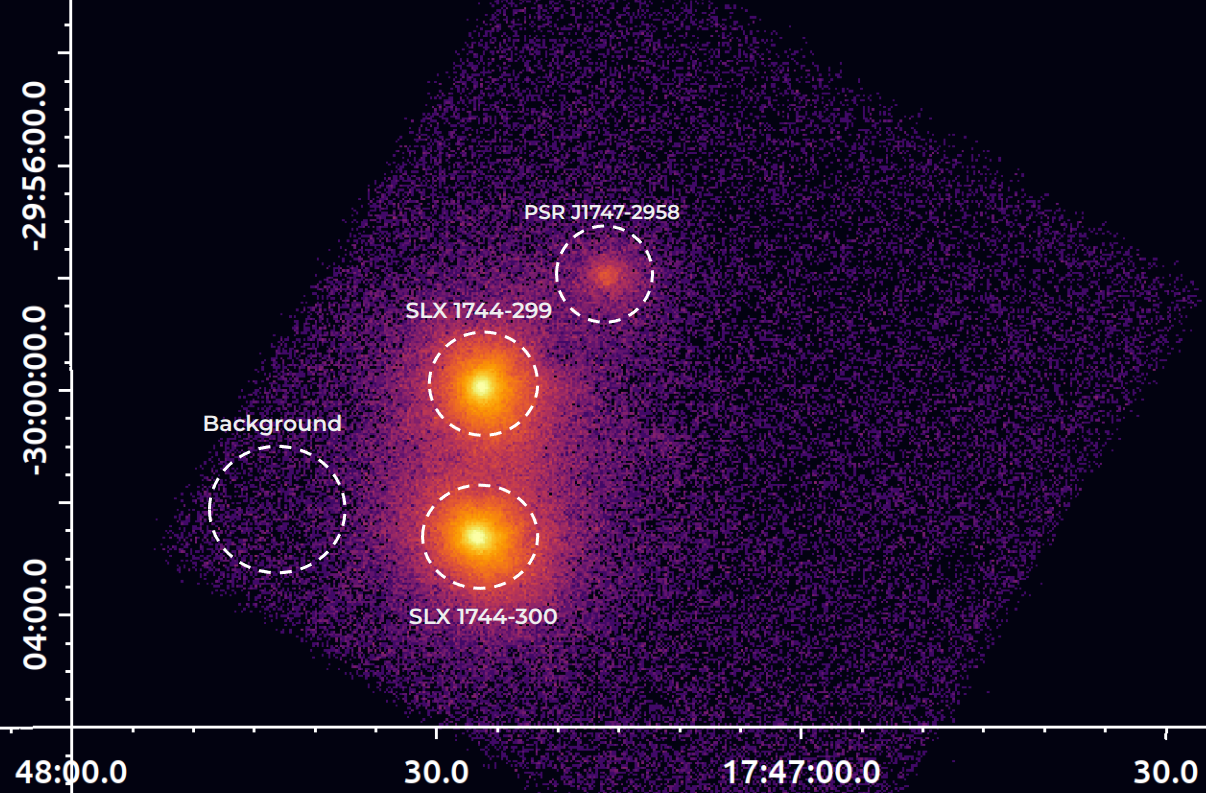}
    \caption{\nustar FPMA module image in the 3--78 keV energy band showing the field of the LMXBs SLX 1744-299 (north) and SLX 1744-300 (south). The pulsar PSR J1747-2958 (also known as ``the Mouse'') is visible within the field. The dashed circular regions indicate the event and background extraction regions for each source.
    }
    \label{Fig:ds9}
\end{figure}

\section{Data reduction and analysis} \label{sec:data}

The {\sl Nuclear Spectroscopic Telescope Array} (\nustar) observed the field on 2018 July 29  \citep[ObsID 30401036002;][]{2013ApJ...770..103H}, with an on-source exposure time of 44.2 ks over an elapsed time of 79 ks. Three sources were detected in this observation: \slxdos, \slxtres and the pulsar PSR~J1747-2958 (see \hyperref[Fig:ds9]{Figure \ref{Fig:ds9}}).

We processed the data using the \nustar Data Analysis Software ({\sc NuSTARDAS}) included in the {\sc HEASOFT} v6.34 package, along with calibration files from CALDB (v.20251215). The {\sc nupipeline} tool employed the parameters {\sc saacalc=1}, {\sc saamode=optimized}, and {\sc tentacle=NO} to filter out passages through the South Atlantic Anomaly (SAA)\footnote{\url{https://nustarsoc.caltech.edu/NuSTAR_Public/NuSTAROperationSite/SAA_Filtering/nulyses_reports/30401036002/nu30401036002_SAA_Report_A.pdf}}. We also verified that no additional correction for the multi-layer insulation tear on the focal plane module A (FPMA) was required \citep{Madsen2020arXiv200500569M}. Barycentric corrections were applied using the clock correction file 20100101v178. Source events were extracted from a circular region with a radius of 50 arcsec, centred on the centroid of the counts distribution for each LMXB. For background events, we used a circular region with a radius of 70 arcsec in a source-free area in the same chip as the targets (see \hyperref[Fig:ds9]{Figure \ref{Fig:ds9}}).

Given the close proximity of the two sources on the sky ($\sim 2.7$ arcmin) and the size of the extraction regions ($R=50$ arcsec), we assessed the potential impact of cross-contamination due to the broad wings of the \nustar\ point spread function. We performed numerical simulations using a King profile, calibrated to match the in-flight full width at-half maximum ($18\arcsec$) and half power diameter ($58\arcsec$; \citealt{2013ApJ...770..103H, Koglin2011SPIE.8147E..0JK}). We estimate cross-contamination fractions of $0.99 \pm 0.07$ per cent for \slxdos\ and $0.76 \pm 0.06$ per cent for \slxtres, and therefore their impact is negligible in our analysis.

Background-subtracted light curves were extracted for both FPMA and FPMB separately and then combined using the {\tt lcmath} task from the FTOOLS package \citep{Blackburn1995ASPC...77..367B}. The extracted spectra from each module were grouped with a minimum of 30 counts per bin, allowing the use of the $\chi^2$ statistic. In the remainder of this section, we describe the temporal and spectral analyses applied to both sources.

\subsection{Temporal analysis}

In order to search for potential pulsation candidates, we analysed the data from the two targets in the 0.1–10 Hz frequency range using the {\tt HENDRICS} tool \citep{hendrics}.
We used the {\sc HENaccelsearch} module to identify optimal pulsation candidates within the specified frequency range. In addition, we derived a 90 per cent confidence upper limit on the pulsed fraction using the {\tt HENz2vspf} tool, based on $10^4$ simulations \citep[see][for more details]{CruzSanchez2026A&A...705A.136C, CruzSanchez2026arXiv260310331C}.
Finally, we generated the Leahy-normalised power density spectra using the {\tt Stingray} software package \citep{stingray2019ApJ...881...39H} and applied the Fourier amplitude difference method to correct for dead time \citep{Bachetti2018ApJ...853L..21B}. We calculated the fractional root-mean-square (rms) amplitude in the 0.1--64 Hz range following the method described by \citet{MunozDarias2011MNRAS.410..679M,MunozDarias2014MNRAS.443.3270M}. Specifically, we used data in the 3--15 keV energy band, divided into 64 s segments. We used a time resolution corresponding to a Nyquist frequency of 1000 Hz; this ensures an accurate estimate of the Poisson (white) noise level when computing the power density spectra.

\subsection{Spectral analysis} \label{subsection:sim}

We simultaneously fitted the 3--78 keV FPMA and FPMB spectra using \textsc{xspec} \citep[v.12.14.0;][]{1996ASPC..101...17A}, with the parameters tied between the two modules. We included a multiplicative constant factor in our models, fixing it to 1.0 for FPMA and allowing it to vary freely for FPMB, to account for differences in their calibration.

We modelled the interstellar absorption using the Tübingen-Boulder model (\textsc{TBabs}), setting the solar abundances according to \citet{2000ApJ...542..914W} and the effective cross-sections to \citet{1996ApJ...465..487V}. We fixed the hydrogen column density ($N_{\rm H}$) to $3.3\times10^{22}\ \mathrm{cm}^{-2}$ for \slxdos and $3.7\times 10^{22}$~cm$^{-2}$ for \slxtres, as obtained by \citet{Mori2005AdSpR..35.1137M} from \xmm observations. Throughout this work, we assumed the upper limits on the distance of $7.2 \pm 1.4\,\mathrm{kpc}$ (SLX~1744$-$299) and $10.3 \pm 0.8\,\mathrm{kpc}$ (SLX~1744$-$300) derived from the analysis of thermonuclear bursts \citep{Chelovekov2017AstL}.

We tested various spectral models, combining up to three components, aimed at addressing the main X-ray spectral features typically observed in NS-LMXBs. For the soft, thermal emission from the accretion disc, we used a multicolour disc (MCD) model \citep[\texttt{diskbb};][]{mitsuda1984PASJ...36..741M, makishima1986ApJ...308..635M}. A blackbody (BB) model (\texttt{bbodyrad}) was used to account for the soft, thermal emission from the NS surface or boundary layer. For the thermally Comptonised emission from the corona, we used the \texttt{nthComp} model, setting the seed-photon geometry (\texttt{inp\_Type}) to 0 for a BB geometry and 1 for an accretion disc geometry \citep{zdziarski1996MNRAS.283..193Z, zycki1999MNRAS.309..561Z}. In our analysis, we fitted the spectra using different combinations of these components. First, we used the \texttt{nthComp} model with seed photon temperature ($kT_{\rm seed}$) as a free parameter.
Additionally, we tested combinations of the \texttt{nthComp} model with an additional soft thermal component (BB or MCD), where $kT_{\rm seed}$ was tied to the temperature of the thermal component, and the seed-photon shape parameter (\texttt{inp\_Type}) was adjusted accordingly to the selected model.

For the BB model, we used the normalisation defined as $R_{\mathrm{km}}^2/D_{10}^2$ ($R_{\mathrm{km}}$ is the source radius in km and $D_{10}$ is the distance to the source in units of 10 kpc) to determine the source radius. For the MCD model, we used the normalisation defined as $(R_{\mathrm{in}}/D_{10})^2 \cos\theta$ (where $R_{\mathrm{in}}$ is the apparent inner disc radius in km and $\theta$ is the inclination angle with respect to the observer, defined as $\theta = 0$ for face-on) to determine the apparent inner disc radius. For the MCD model we tested inclination angles of $40^\circ$ and $70^\circ$, applied a correction factor for the torque-free inner boundary condition ($\xi=0.4$) and the ratio of colour temperature to effective temperature \citep[$\kappa=1.7$;][]{Kubota1998PASJ...50..667K}. The reported fluxes were calculated using the convolutional model {\tt cflux}. 

To assess the possible presence of the Fe K$\alpha$ emission line at $\sim 6.4$ keV, we performed over $10^{5}$ spectral simulations in \textsc{xspec} to determine the significance level for detecting this feature. We used the observational dataset with the {\sc fakeit} command and generated random parameters with the {\sc simpars} command.  By constructing a cumulative distribution function of F-values from the simulations, we established the minimum significance level for detection by comparing the simulated F-values to the F-value from the data. We defined the null hypothesis as the continuum-only model (in \textsc{xspec} [{\sc const${\times}$TBabs${\times}$(nthComp)}]), while the tested hypothesis included both the continuum and the Gaussian component ({\sc gauss}). We determined the significance level by calculating the associated $p$ value, representing the fraction of simulated spectra with F-values greater than those obtained by fitting the real data. This simulation-based approach controls the false-positive rate when testing for an additional Gaussian component.

To account for \nustar\ calibration uncertainties \citep{Madsen2015ApJS..220....8M, Grefenstette2022arXiv220604058G, Madsen2022JATIS...8c4003M}, we allowed the FPMA gain offset to vary while keeping the FPMB gain fixed. We obtained best-fitting offsets of $-47$ eV for SLX 1744-299 and $-72$ eV for SLX 1744-300, consistent with the typical $\sim$40--80 eV systematic uncertainties reported in the literature \citep[e.g.][]{Zalot2024A&A...686A..95Z, Ballhausen2024A&A...688A.214B, LaMonca2024ApJ...960L..11L, Diez2023A&A...674A.147D}.

We calculated the parameter errors at the 90 per cent confidence level using the Markov chain Monte Carlo technique implemented in \textsc{xspec}.  We used the Goodman-Weare algorithm with a total of $10^7$ steps, with walkers set to 10 times the number of free parameters \citep{Goodman2010CAMCS...5...65G}. We verified convergence by inspecting the parameters' trace plots and by checking that the estimated autocorrelation times were stable (see \citealt{saavedragx13, Saavedra2023A&A...680A..88S} for more details).

 \begin{figure*}
    \centering
    \includegraphics[width=\columnwidth]{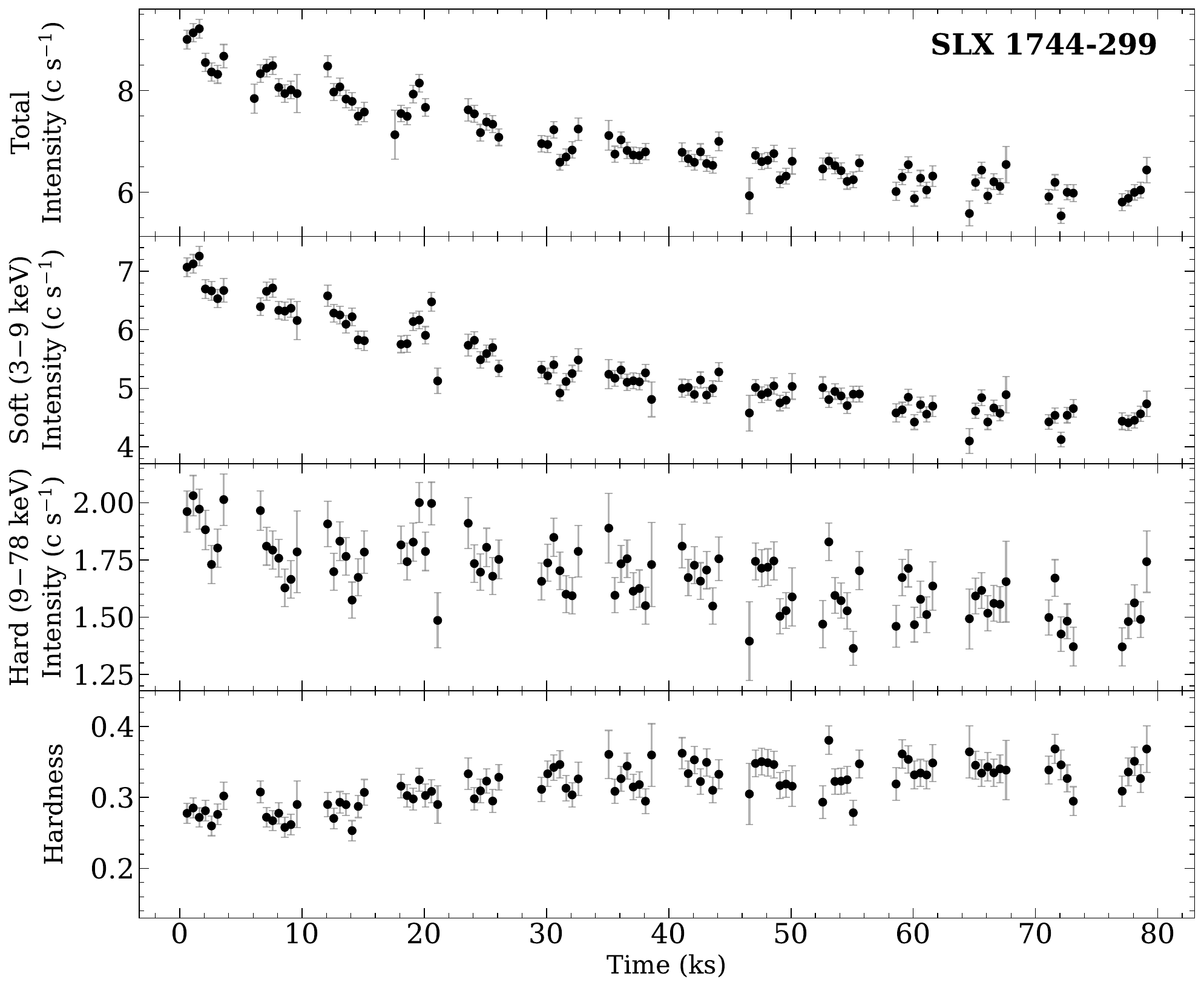}
    \includegraphics[width=\columnwidth]{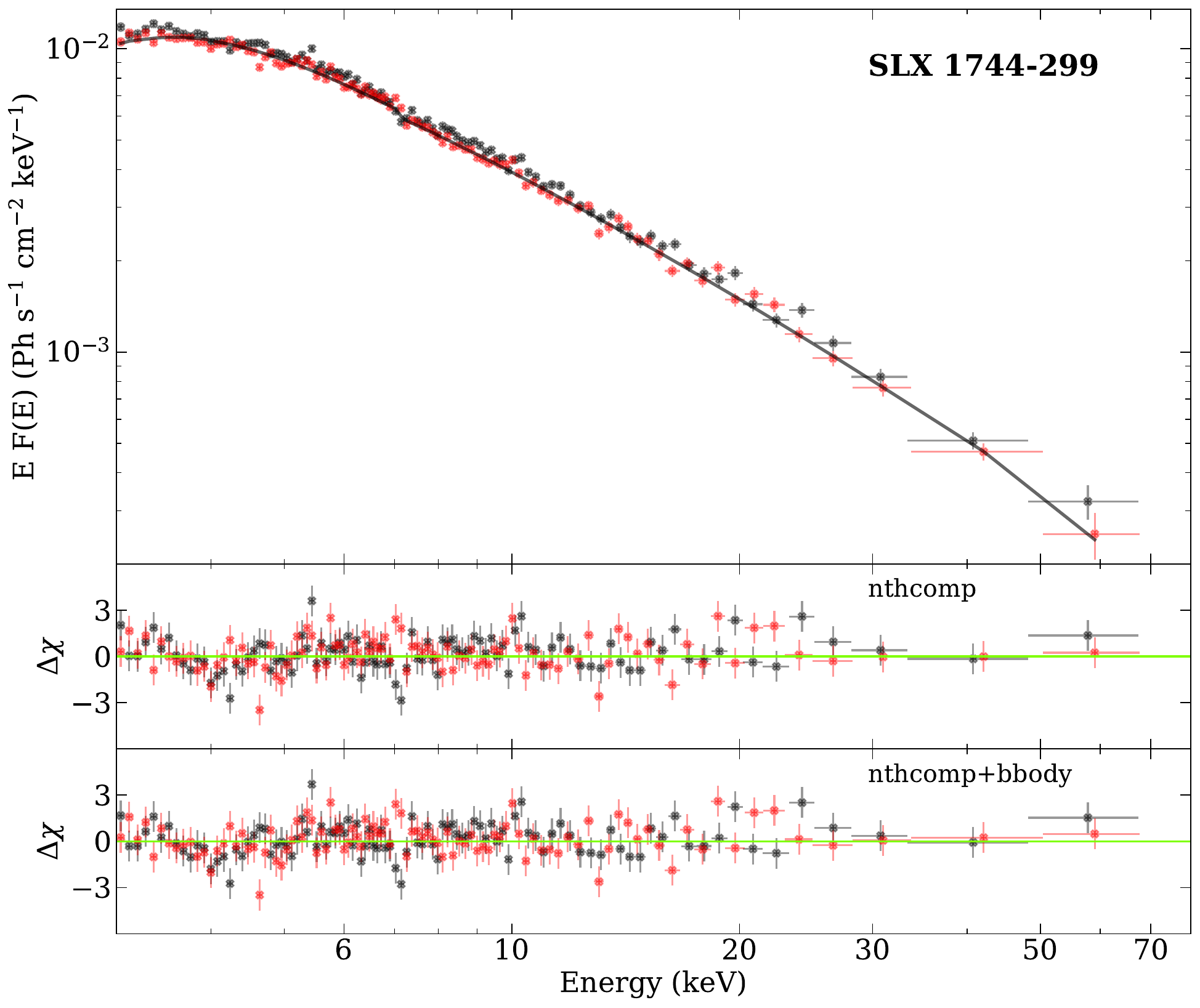}
    \caption{{\it Left panel}: Background-corrected \textit{NuSTAR} light curve of SLX 1744$-$299 with a time bin of 500 sec. From top to bottom: the total light curve in the 3--78 keV energy range, the soft band (3--9 keV), the hard band (9--78 keV), and the hardness ratio (hard-to-soft) as a function of time. {\it Right panel}: Unfolded \textit{NuSTAR} FPMA (red) and FPMB (black) spectrum (top) and residuals (middle and bottom) using the \texttt{const${\times}$TBabs${\times}$(nthComp)} model. The bottom panel shows the fit residuals when using \texttt{const${\times}$TBabs${\times}$(bbodyrad+nthComp)}.
    }
    \label{Fig:slx299}
\end{figure*}

\begin{figure} [h!]
\centering
    \includegraphics[width=0.9\columnwidth]{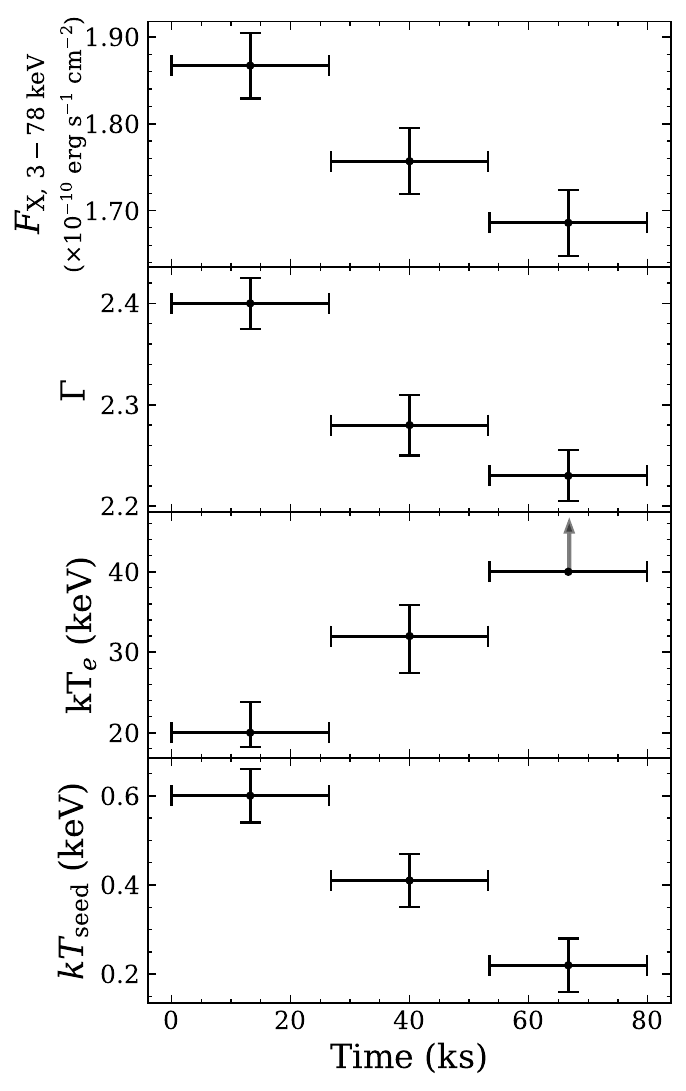}
    \caption{Time-resolved spectral evolution of SLX~1744$-$299. From top to bottom: unabsorbed 3--78~keV X-ray flux, photon index
($\Gamma$), electron temperature ($kT_{\mathrm{e}}$), and seed-photon temperature ($kT_{\mathrm{seed}}$). Error bars represent 90 per cent confidence intervals.
}
    \label{fig:resolved_compt}
\end{figure}

\section{\slxdos: Results} \label{sec:result299}

\subsection{Time-averaged properties}

Our detailed temporal analysis of \slxdos\ did not reveal any coherent signals, with a pulsed fraction upper limit of 3.9 per cent (90 per cent confidence), but the source exhibited significant aperiodic variability. From the 3--15~keV light curve, we measured a fractional rms variability of $23.2 \pm 3.1$~per cent (1$\sigma$), consistent with values typically observed in NS-LMXBs during the hard state \citep{MunozDarias2014MNRAS.443.3270M}, indicating that \slxdos\ was in this accretion regime during the \nustar\ observation.

We modelled the spectra using a standard approach for NS-LMXBs in the hard state. Our initial fit used a single thermally Comptonised continuum model affected by photoelectric absorption, assuming two scenarios: seed photons originating either from a BB or from a MCD (i.e. setting \texttt{inp\_type} to 0 and 1, respectively). The two assumptions provided equally good statistical fits ($\chi^{2}_{\nu}$ = 1.04 for 856 dof; p-value = 0.20), with no evident structured residuals across the energy range, including the Fe K$\alpha$ region (Fig.~\ref{Fig:slx299}, right panel). To quantify this, we tested for the presence of a narrow Fe emission line by adding a Gaussian component, but it was not significantly detected ($2.1\sigma$; see Sect.~\ref{subsection:sim}), and was therefore not included in any of the spectral models. For the BB scenario, we obtained $kT_{\rm seed} = 0.58\pm0.02$ keV; in the case of the MCD, we found $kT_{\rm seed} = 0.7\pm0.03$ keV. In both cases, we obtained a $\Gamma = 2.33\pm0.01$ and a $kT_e = 38^{+53}_{-12}$ keV. The unabsorbed 3--78~keV flux is $(1.77\pm0.01) \times 10^{-10}~\erg\,\s^{-1}\,\cm^{-2}$, which corresponds to an upper limit on the luminosity of $(1.06\pm0.51)\times 10^{36}~\erg\,\s^{-1}$.

Adding a BB component (\texttt{bbodyrad}) to the model did not significantly improve the fit ($\Delta\chi^2 = 0.6$ for 1 dof; F-test probability of 0.47). The model parameters remained consistent with the previous fit, including the temperature of the BB component (which corresponds to the seed photon temperature, $kT_{\rm bb} = kT_{\rm seed}$). From the BB normalisation, we derived an emission radius of $R_{\rm bb} < 1.3$~km. In this model, the Comptonisation component accounts for $> 98$ per cent of the total 3--78~keV unabsorbed flux, while the thermal contribution is $< 2$ per cent. \hyperref[tab:params]{Table~\ref{tab:params}} presents the detailed results of the fitted models, and \hyperref[Fig:slx299]{Fig.~\ref{Fig:slx299}} (right panel) displays the corresponding spectrum and residuals.

We also tested a model consisting of an MCD plus Comptonisation. The fit returned parameters consistent with those from the single \texttt{nthComp} model (e.g. ${kT_{\rm in} =0.71\pm0.03}$~keV). However, the corrected inner disc radius inferred from the {\tt diskbb} normalisation ($N_{\rm diskbb} < 1.2 \times 10^{-5}$) corresponds to $R_{\rm in} < 3$~m, which is unphysically small compared to the typical NS radius of 10~km. Therefore, we discarded the model.

\subsection{Time-resolved spectral analysis}
\label{spectral_299}

The \nustar\ light curve of SLX~1744$-$299 (3--78~keV, 500~s bins) shows a gradual decline in count rate from ${9.0\pm0.2}$ to ${6.4\pm0.2}$~cts~s$^{-1}$, corresponding to a ${28}$ per cent decrease, over the 79 ks observation (see \hyperref[Fig:slx299]{Fig.~\ref{Fig:slx299}}, top panel). When inspecting the light curve in different energy bands, the drop appears more pronounced in the soft range (3--9~keV), while the hard range (9--78~keV) remains relatively stable. Consequently, the hardness ratio increases from $0.28\pm0.01$ to $0.36\pm0.02$ during the first $40$~ks. 

In order to investigate the physical origin of this spectral evolution, we performed a time-resolved spectral analysis. We divided the observation into three consecutive 26 ks intervals and extracted one spectrum per segment. We fitted these three spectra using the preferred model identified for the entire observation (i.e., \texttt{const${\times}$TBabs${\times}$(nthComp)} with a BB geometry for the seed photons). 

The model fitted all three intervals acceptably, yielding reduced chi-squared values (chi-squared and degrees of freedom) of 1.05 (733/692), 1.05 (721/686), and 1.01 (608/598), respectively. Our analysis revealed three key trends (see \hyperref[fig:resolved_compt]{Fig. \ref{fig:resolved_compt}}): first, a slight hardening of the photon index, which decreased from $2.47\pm0.03$ to $2.26\pm0.03$; second, a steady decline in the seed-photon temperature from $0.60\pm0.02$ keV to $0.21\pm0.03$ keV; and third, an increase in the electron temperature from $20_{-2}^{+4}$ keV to $\gtrsim40$ keV (the latter being a lower limit).  

This combination of a hardening photon index, a cooler seed photon temperature, and a hotter electron temperature is physically consistent with a declining mass-accretion rate ($\dot{M}$). This scenario suggests that the corona becomes less efficiently cooled as the observation progresses (see Sec.~\ref{subsec:variability}).

\begin{table*}[ht]
\centering
\caption{
Best-fit spectral results for \slxdos\ and \slxtres.
}
\resizebox{2\columnwidth}{!}{%
\renewcommand{\arraystretch}{1.4} 
\begin{tabular}{cccccc}
    \toprule
    Component   & Parameter & \multicolumn{2}{c}{\slxdos} & \multicolumn{2}{c}{\slxtres} \\ 
    & & {\sc nthComp} & {\sc nthComp+bbodyrad} & {\sc nthComp} & {\sc nthComp+bbodyrad} \\
    \midrule
    {\sc const} & $C_{\rm AB}$ & $1.03\pm0.01$& $1.03\pm0.01$& $1.03\pm0.01$& $1.03\pm0.01$\\
    \midrule
    {\sc TBabs} & $N_{\rm H}$ (10$^{22}$~cm$^{-2}$) & $3.3^\dagger$ & $3.3^\dagger$ & $3.7^\dagger$ & $3.7^\dagger$ \\          
    \midrule
    {\sc nthComp} & $\Gamma$ & $2.33\pm0.01$ & $2.32\pm0.01$& $2.28\pm0.02$& $2.28\pm0.02$\\
    & $kT_{\rm e}$ (keV) & $38_{-12}^{+53}$& $37_{-12}^{+19}$&  $8.7\pm0.8$&  $8.7\pm0.5$\\
    & $kT_{\rm seed}$ (keV) & $0.58\pm0.02$& $=kT_{\rm bb}$ & $0.63\pm0.03$& $=kT_{\rm bb}$  \\
    & norm$_{\rm compt}$ ($\times 10^{-2}$) & $1.3\pm0.1$& $1.3\pm0.1$& $1.4\pm0.1$& $1.4\pm0.1$\\
    \midrule  
    {\sc bbodyrad} & $kT_{\rm bb}$ (keV) & -  & $0.57\pm0.04$& - & $0.64\pm0.03$\\
    & norm$_{\rm bb}$  & - & $<3.3$& - & $<1.4$\\
    & $R_{\rm bb}$ [km]~$^b$ & - & $<1.3$& -  & $<1.2$ \\
    \midrule  
    \multicolumn{2}{c}{{\sc ${\it F}_{\rm X}^{a}$} [0.5--10 keV]} & $1.49\pm0.01$ & $1.54\pm0.01$ &  $1.93\pm0.01$ &  $1.93\pm0.01$  \\ 
    \multicolumn{2}{c}{{\sc ${\it F}_{\rm X}^{a}$} [2--10 keV]} & $1.33\pm0.01$ & $1.36\pm0.01$ &  $1.73\pm0.01$ &  $1.73\pm0.01$  \\ 
    \multicolumn{2}{c}{{\sc ${\it F}_{\rm X}^{a}$} [3--78 keV]} & $1.77\pm0.01$ & $1.77\pm0.01$ & $2.02\pm0.01$ & $2.02\pm0.01$ \\ 
    \midrule    
    \multicolumn{2}{c}{{\sc ${\it L}_{\rm X}^{b}$} [10$^{36}~\lum$]}  & $<1.06\pm0.51$ & $<1.06\pm0.51$ & $<2.56\pm0.43$ & $<2.56\pm0.43$  \\ 
    \bottomrule
    \multicolumn{2}{c}{$\chi^{2}_{\nu}/{\rm dof}$} & 1.04/856 & 1.04/855 & 1.03/797 & 1.03/796 \\
    \bottomrule
\end{tabular}%
}
\tablefoot{We used a thermal Comptonisation model, \texttt{const${\times}$TBabs${\times}$(nthComp)}, with \texttt{inp\_Type = 0} and free $kT_{\rm seed}$. We also used a combined model, \texttt{const${\times}$TBabs${\times}$(nthComp+bbodyrad)}, in which $kT_{\rm seed}$ is tied to the BB temperature ($kT_{\rm bb}$). All parameter uncertainties are reported at the 90 per cent confidence level. $^\dagger$ Corresponding parameter frozen during the fit. $^a$ Unabsorbed X-ray flux in units of 10$^{-10}~\flux$. $^b$ Calculated in the 3--78 keV range, assuming an upper limit of $7.2\pm1.4$ kpc for \slxdos\ and $10.3\pm0.8$ kpc for \slxtres.}
\label{tab:params}
\end{table*}

 \begin{figure*} [h!]
\centering
    \includegraphics[width=\columnwidth]{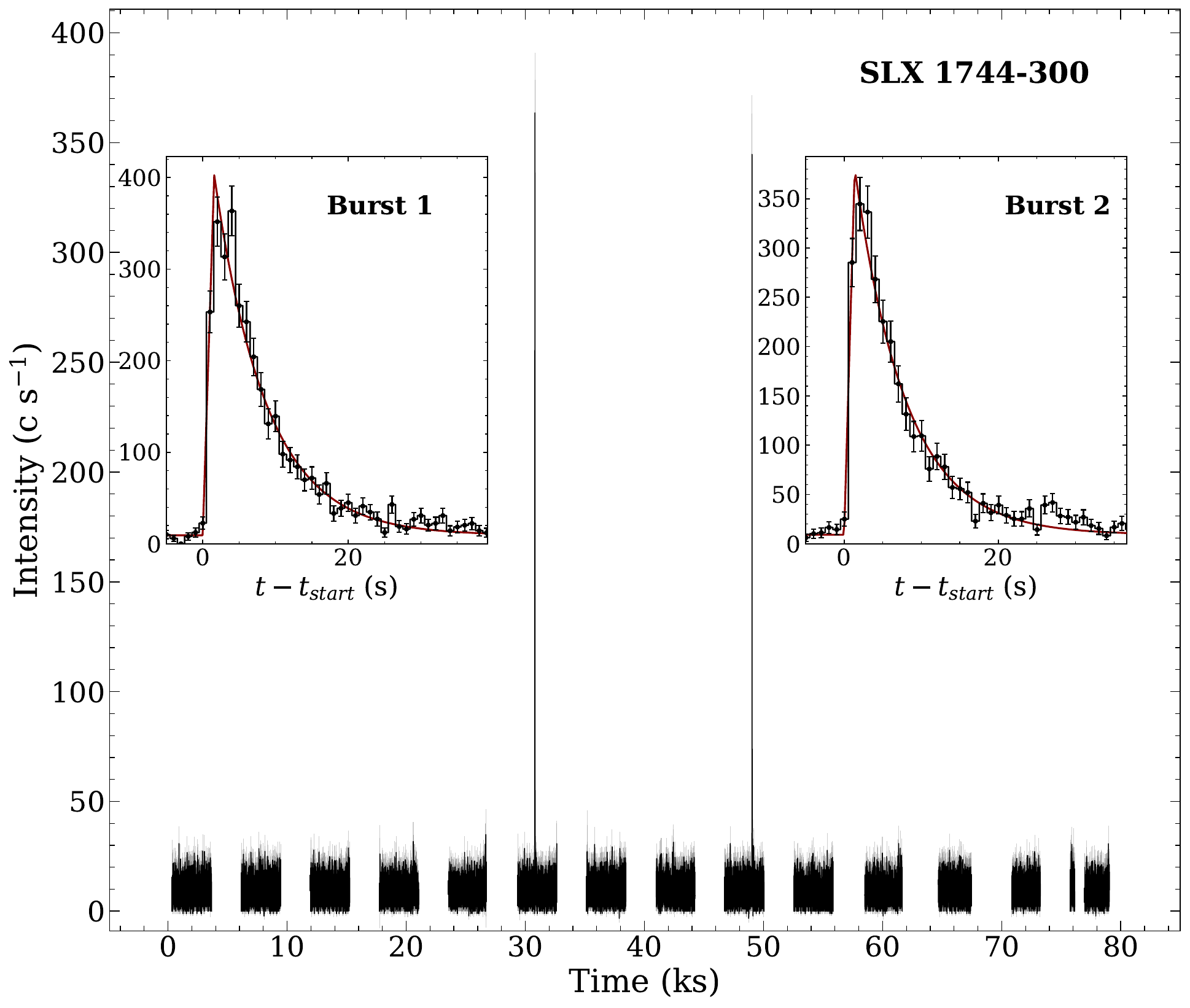}
    \includegraphics[width=\columnwidth]{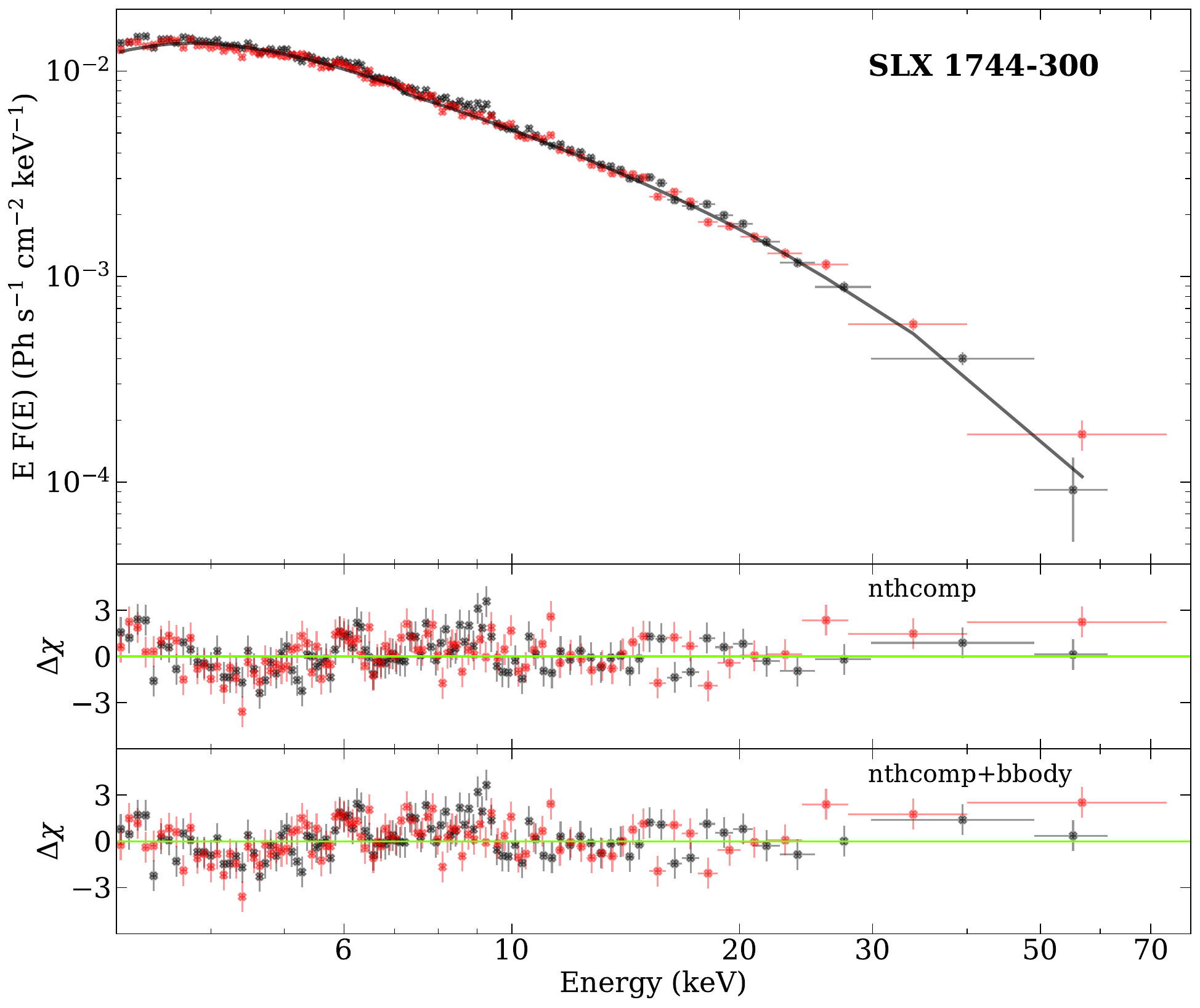}
    \caption{{\it Left panel}: Background-corrected \textit{NuSTAR} light curve of SLX 1744$-$300 with a time bin of 1~s, including two Type-I bursts superimposed on the persistent emission. The insets show each burst and the best-fit model from Eq.~\ref{eq:burst}.
    {\it Right panel}: Unfolded \textit{NuSTAR} FPMA (red) and FPMB (black) spectrum (top) and residuals (middle) using the \texttt{const${\times}$TBabs${\times}$(nthComp)} model. The bottom panel shows the fit residuals when using \texttt{const${\times}$TBabs${\times}$(bbodyrad+nthComp)}.}
    \label{fig:slx300}
\end{figure*}

\section{\slxtres: Results}  \label{sec:result300}

The \nustar\ light curve of \slxtres\ revealed two prominent Type-I bursts (see the top panel of \hyperref[fig:slx300]{Fig.~\ref{fig:slx300}}). These events were excluded from the persistent emission analysis and are examined separately in Section~\ref{sec:burst_analysis_300}. In the following, we first characterise the persistent emission and then analyse the bursts.

\subsection{Persistent X-ray emission properties}

We searched for periodic signals but found no significant periodicities, with a pulsed fraction upper limit of 9.1 per cent (90 per cent confidence). We determined the fractional rms variability to be $22.6 \pm 4.3$ per cent ($1\sigma$), which indicated that the source was in the hard state at the time of the observation \citep{MunozDarias2014MNRAS.443.3270M}.

We fitted the time-averaged spectra of the persistent emission with an absorbed thermal Comptonisation model. As in the case of \slxdos\ (Section~\ref{sec:result299}), we considered two configurations for the seed photon source: the BB and the MCD. In both cases, the fits were statistically acceptable ($\chi^{2}_{\nu} = 1.03$ for 797 dof; p-value = 0.27). No significant Fe K$\alpha$ emission line was detected ($2.8\sigma$; see Sect.~\ref{subsection:sim}). The BB scenario yielded a seed temperature $kT_{\rm seed} = 0.63\pm0.03$ keV, whereas the MCD case resulted in $kT_{\rm seed} = 0.76\pm0.04$ keV. The two assumptions returned consistent values for the photon index ($\Gamma = 2.28 \pm 0.02$) and the electron temperature ($kT_{\rm e} = 8.7 \pm 0.8~$ keV). The unabsorbed 3--78~keV flux is $(2.02\pm0.01) \times 10^{-10}~\flux$, which corresponds to $L_{\rm X} < (2.56 \pm 0.43) \times 10^{36}~\lum$.

Incorporating an additional {\tt bbodyrad} component into the model did not significantly improve the fit ($\Delta\chi^2=0.8$ for 1 dof; F-test probability of 0.87). Using the normalisation from this component and adopting a distance of $10.3 \pm 0.8$ kpc, we estimated an upper limit of 1.2 km for the radius of this potential thermal emission region. Within this model, the Comptonisation component contributes $> 99$ per cent of the unabsorbed 3--78~keV flux, while the thermal fraction is $< 1$ per cent. \hyperref[tab:params]{Table~\ref{tab:params}} presents the detailed parameters, and \hyperref[fig:slx300]{Fig.~\ref{fig:slx300}} (right panel) illustrates the spectrum and model residuals.

Similarly to the case of \slxdos, the MCD plus Comptonisation model produced unphysical results. This model returned a seed photon temperature ${kT_{\rm in} = 0.76\pm0.03}$ keV. The corrected inner disc radius, which we inferred from the \texttt{diskbb} normalisation ($N_{\rm diskbb} < 5.6 \times 10^{-5}$), was unphysically small ($R_{\rm diskbb} < 11~\m$ for a $40^\circ$ inclination and $< 16~\m$ for $70^\circ$). We therefore did not consider this model further.

\subsection{Type-I thermonuclear burst properties}
\label{sec:burst_analysis_300}

\slxtres\ exhibited two thermonuclear Type-I bursts during the \nustar\ observation. To characterise their temporal profiles, we fitted the burst profiles using the QDP {\sc burs} model\footnote{See \url{https://heasarc.gsfc.nasa.gov/docs/software/ftools/others/qdp/node143.html}}. This model represents each burst as a linear rise followed by an exponential decay, superimposed on a constant representing the persistent emission level. The model is defined as:

\begin{equation}
R(t) =
\begin{cases}
    F_{\rm pers}  & t < t_{\rm start} \\
    F_{\rm pers} + A \cdot \frac{(t - t_{\rm start})}{t_{\rm peak} - t_{\rm start}} & t_{\rm start} < t < t_{\rm peak} \\
    F_{\rm pers} + A \cdot e^{-(t - t_{\rm peak})/D_T} &  t_{\rm peak} < t,
\end{cases}
\label{eq:burst}
\end{equation}

\noindent where ${F}_{\rm pers}$ is the rate associated with the persistent emission. ${t_{\rm start}}$ and ${t_{\rm peak}}$ are the burst start time and peak time, respectively, indicating the onset and maximum intensity moments of the burst. The parameter ${A}$ is the burst amplitude above the persistent level, and ${D_T}$ is the decay time constant, which controls the rate at which the burst count rate decreases after its peak. The two bursts are shown at 1-s resolution in the insets of  Fig.~\ref{fig:slx300}, together with the best-fit model from  Eq.~\ref{eq:burst}. We implemented a Monte Carlo sampling method to determine the $1\sigma$ uncertainties of the parameters, which are summarised in \hyperref[tab:burst]{Table~\ref{tab:burst}}. 

Burst-1 started at $30804.2_{-0.1}^{+0.5}$ s and lasted $34.2\pm1.2$ s. The rise time was $1.5_{-0.6}^{+0.3}$~s, and the decay time was $7.2\pm0.3$ s. The emission preceding Burst-1 was measured at $9.3\pm0.8$~cts~s$^{-1}$, reaching a peak of $402.5$~cts~s$^{-1}$. Burst-2 started at $49033.2_{-0.1}^{+0.4}$~s and lasted $31.6\pm1.4$~s. The rise and decay times were $1.4_{-0.5}^{+0.2}$~s and $6.6\pm0.3$~s, respectively. The emission preceding Burst-2 was measured at $9.2\pm0.7$~cts~s$^{-1}$, reaching a peak of $380.2$~cts~s$^{-1}$. The bursts are consistent with mixed H and He fuel (see Section~\ref{subsec:Discburst}).

\renewcommand{\arraystretch}{1.5}
\begin{table*}[h!]
\caption{Best-fit burst parameters.}
\resizebox{\textwidth}{!}{%
\begin{tabular}{@{}ccccccc@{}}
\toprule
Burst No & \begin{tabular}[c]{@{}c@{}}Burst start time\\ (s)\end{tabular} & \begin{tabular}[c]{@{}c@{}}Burst duration\\ (s)\end{tabular} & \begin{tabular}[c]{@{}c@{}}Rise time\\ (s)\end{tabular} & \begin{tabular}[c]{@{}c@{}}Decay time\\ (s)\end{tabular} & \begin{tabular}[c]{@{}c@{}}Persistent emission \\ (cts s$^{-1}$)\end{tabular} & \begin{tabular}[c]{@{}c@{}}Peak count rate\\ (cts s$^{-1}$)\end{tabular} \\ \midrule
$1$ & $30804.2_{-0.1}^{+0.5}$ & $34.2\pm1.2$ & $1.5_{-0.6}^{+0.3}$ & $7.2\pm0.3$ & $9.3\pm0.8$ & $402.5$ \\
$2$ & $49033.2_{-0.1}^{+0.4}$ & $31.6\pm1.4$ & $1.4_{-0.5}^{+0.2}$ & $6.6\pm0.3$ & $9.2\pm0.7$ & $380.2$ \\ \bottomrule
\end{tabular}%
}
\tablefoot{The two type-I X-ray bursts were modelled with the {\sc burs} model. Uncertainties correspond to $1\sigma$.}
\label{tab:burst}
\end{table*}

To investigate the spectral evolution during these events, we also conducted a time-resolved spectroscopic study. We fitted the spectra from each burst using a fixed model for the persistent emission (i.e. a single thermal Comptonisation continuum), while allowing a soft thermal BB component to vary \citep[e.g.,][]{intZand2011A&A...525A.111I, Degenaar2016MNRAS.456.4256D}. By exploring different good time intervals, we achieved an optimal temporal resolution of six spectra per burst.

This analysis showed spectral cooling in both bursts. For burst-1, the thermal temperature cooled from $2.61\pm0.35$ to $1.42\pm0.21~\mathrm{keV}$, while the apparent emission radius inferred from the BB normalisation increased from $3.1_{-0.9}^{+1.4}$ to $6.1_{-1.6}^{+2.0}~\mathrm{km}$, assuming $d = 10.3\,\mathrm{kpc}$. For burst-2, the temperature cooled from $1.85\pm0.24$ to $0.98\pm0.15~\mathrm{keV}$, while the apparent radius increased from $4.1_{-1.1}^{+1.3}$ to $7.7_{-2.8}^{+4.2}~\mathrm{km}$.
The corresponding bolometric burst fluences ($f_b$), derived from the time-resolved unabsorbed 0.01--100~keV fluxes, were $(6.4 \pm 0.4) \times 10^{-8}~\fb$ and $(5.1 \pm 0.4) \times 10^{-8}~\fb$  for the first and second bursts, respectively.
We found no evidence of photospheric radius expansion in either event, which prevented us from further constraining the source distance.

\section{Discussion}
\label{sec:discussion}

Low-mass X-ray binaries at X-ray luminosities $\lesssim 0.01 L_{\rm Edd}$ provide crucial insights into accretion physics, although their study is often limited by observational constraints. The two LMXBs, \slxdos\ and \slxtres\, offer an excellent opportunity to probe this low-luminosity regime, as they are persistent systems accreting at $\lesssim 0.01~L_{\rm Edd}$. However, their small angular separation has historically made it difficult to disentangle their individual contributions. For the first time above 10~keV, the exceptional spatial resolution and broad 3--78~keV energy coverage of \nustar\ allowed us to carry out a detailed spectral and timing study of these two sources.

\subsection{The low-luminosity hard state}

From the \nustar\ observation, we were able, for the first time, to measure the integrated fractional rms (0.1–64 Hz) for \slxdos\ and \slxtres\ individually. We obtained values of $\sim$~21–-23 per cent in both cases, consistent with the sources being in the hard state during the observation \citep{MunozDarias2014MNRAS.443.3270M}. The spectral analysis of the systems is consistent with the above picture. The X-ray spectra of both sources are well described by a simple absorbed thermally Comptonised model.

The spectral fitting returned similar values of the photon indices ($\Gamma \sim 2.3$) and seed photon temperatures ($kT_{\rm seed}\sim0.6$ keV), assuming the seed photons arise from the NS surface or boundary layer. However, the electron temperature of the Comptonising corona differed between the two sources: for \slxdos\ we found $kT_{\rm e} \approx 38$ keV, while for \slxtres\ it was $kT_{\rm e} \approx 9$ keV. These $\Gamma$ and $kT_{\rm e}$ values translated to an optical depth\footnote{The electron scattering optical depth ($\tau$) is obtained following the relation $\Gamma_{\tau} = \bigl[9/4 + \bigl((kT_{\rm e}/m_{\rm e}c^2)\,\tau\,(1+\tau/3)\bigr)^{-1}\bigr]^{1/2} - 1/2$.} of $\tau\approx1.5$ and $\approx4.4$ for \slxdos\ and \slxtres, respectively. During hard state, NS-LMXBs show higher electron temperatures ($kT_{\rm e} \gtrsim 10 $ keV) and lower optical depths ($\tau \lesssim 3$) than in the soft state (where $kT_{\rm e} \lesssim 5$ keV and $\tau \gtrsim 5$ is typically found; see e.g. \citealt{Lin2007ApJ...667.1073L, ArmasPadilla2017MNRAS.467..290A, Burke2017MNRAS.466..194B}). Hence, \slxdos is fully consistent with typical hard state values while \slxtres shows a slightly higher $\tau$ than typical. 

To test whether the thermal emission originates from the NS surface or boundary layer, or from the inner disc, we added either BB or MCD components to the Comptonisation model. We found that the data (i.e. the {\it NuSTAR} band) did not statistically require any additional thermal component. Including a BB model, the parameters of the Comptonisation component remained unchanged. The BB temperature was consistent with the previous seed photon temperature ($kT_{\rm bb} \approx 0.6$ keV), and the component contributed approximately $<2$ per cent and $<1$ per cent to the 3--78 keV flux for \slxdos\ and \slxtres, respectively.
When we replaced the BB with a MCD model, the fit returned unphysically small inner-disc radii. 

We note that the {\it NuSTAR} spectral coverage starts at 3 keV and thus offers limited sensitivity to very soft thermal emission below that threshold. Nonetheless, the absence of a significant soft component in our spectra is consistent with established luminosity-dependent trends for NS-LMXBs: soft thermal components typically contribute $\sim$30--50 per cent of the 0.5--10 keV emission at low luminosities $L_{0.5-10}\lesssim10^{35}\ \mathrm{erg\ s^{-1}}$ ($\sim20$ per cent in the 0.8–30 keV band), attributed to residual accretion on the NS surface \citep{zamperi1995ApJ...439..849Z,armaspadilla3030_2013MNRAS.436L..89A,Baharamian2017MNRAS.467.2199B,Degennar2013ApJ...767L..31D,ArmasPadilla2013MNRAS.434.1586A,Arnason2015ApJ...807...52A,ArmasPadilla2018MNRAS.473.3789A, ArmasPadilla2017MNRAS.467..290A}. Above $10^{35}\ \mathrm{erg\ s^{-1}}$, however, this component weakens or, in some cases, disappears \citep{ArmasPadilla2013MNRAS.434.1586A,Allen2015ApJ...801...10A,Wijnands2015MNRAS.454.1371W, Stoop2021MNRAS.507..330S}. In this context, several broad-band studies have shown that the dominant component is thermal Comptonisation: the soft thermal component typically contributes only $\lesssim10$–20 per cent at energies up to $\sim80$~keV, while the Comptonised continuum carries most of the broad-band emission \citep[e.g.][]{Degenaar2015MNRAS.451L..85D, Matranga2017A&A...603A..39M, Ludlam2017ApJ...836..140L, vandenEijnden2018MNRAS.475.2027V}. Therefore, our estimated luminosities ($L_{\rm X} \lesssim 1.06 \times 10^{36}~\lum$ for \slxdos\ and $L_{\rm X} \lesssim 2.56 \times 10^{36}~\lum$ for \slxtres) are consistent with hard-state NS-LMXBs in which any soft thermal component contributes only a minor fraction of the 3--78 keV flux and remains undetectable in the \nustar\ band.

\subsection{Variability across different timescales}
\label{subsec:variability}

The small angular separation between \slxtres and \slxdos ($\sim2.7$ arcmin) has long hindered efforts to measure their individual X-ray flux contributions. This limitation was first overcome by \citet{Mori2005AdSpR..35.1137M}, who used an \xmm observation to spatially resolve the two sources and report their respective 0.5--10 keV fluxes. In that work, the absorbed 0.5--10 keV flux of \slxdos was reported to be higher than that of \slxtres by a factor of 1.9, with measured absorbed fluxes of $1.9\times10^{-10}~\flux$ and $1.0\times10^{-10}~\flux$, respectively.
In contrast, our \nustar\ observation provides, for the first time above 10~keV, spatially resolved flux measurements over the 3--78~keV energy band. We measured unabsorbed fluxes of $1.8\times10^{-10}~\flux$ and $2\times10^{-10}~\flux$ for \slxdos\ and \slxtres, respectively, implying that \slxtres\ is now slightly brighter, with a flux ratio of $\sim1.15$ ($\sim1.3$ when extrapolated to the 0.5--10~keV band).
We note that long-term monitoring of the field by the \textit{Rossi X-ray Timing Explorer} \citep{intZand2007A&A...465..953I} also showed significant variability. In addition, our observations show variability on timescales of a few tens of kiloseconds. All the above suggests that the flux ratio between the two sources is not fixed and that it changes across short and long timescales.

\subsubsection{\slxdos}

We performed time-resolved spectroscopy on \slxdos\ to trace its spectral evolution throughout the observation. Our analysis revealed a gradual flux decline of 28 per cent over the 79 ks exposure. As the flux decreased, the spectrum hardened: the photon index decreased from $\sim2.4$ to $\sim2.2$, the seed-photon temperature cooled from $\sim0.6$ to $\sim0.2$\,keV, and the coronal temperature rose from $\sim20$ to $\gtrsim40$\,keV. This spectral evolution is consistent with a decreasing mass-accretion rate. A lower accretion rate supplied fewer seed photons from the boundary layer or the inner accretion disc, making inverse-Compton cooling less efficient. This reduced cooling allowed the coronal electrons to heat up, increasing $kT_{\mathrm{e}}$ and hardening the emitted spectrum -- naturally explaining the observed evolution \citep[see e.g.][]{Done2007A&ARv..15....1D}.

This steady decrease in source intensity is reminiscent of the behaviour observed in other NS-LMXBs. In particular, similar trends have been reported for the transient UCXB IGR~J17494$-$3030 \citep{armaspadilla3030_2013MNRAS.436L..89A} and the UCXB candidate XTE~J1709$-$267 \citep{Degennar2013ApJ...767L..31D}. In both systems, \xmm\ observations revealed a comparable decline in flux at $L_{\rm X}< 10^{35}~\lum$. However, unlike our case, time-resolved spectroscopy showed that the decrease in intensity was accompanied by a drop in the inferred temperature of the thermal component, while the photon index ($\Gamma$) remained constant. In those works, the flux decay was interpreted as the result of decreasing low-level accretion onto the NS surface, which produced the observed soft thermal emission (we note that although crustal cooling was initially suggested for XTE~J1709$-$267, it was later disfavoured; \citealt{armaspadilla3030_2013MNRAS.436L..89A}).

In our case, the \nustar\ low-energy limit (3~keV) prevents us from constraining the soft thermal component, and therefore from confirming or ruling out whether a similar trend is taking place. Still, we observe a decrease in the seed photon temperature, which we assume originates from the NS surface, suggesting a potentially analogous scenario. However, we also detect a clear spectral evolution in the Comptonisation component, including an increase in the electron temperature and a hardening of the photon index. These results suggest an overall spectral evolution, encompassing both seed photon and Comptonisation properties, suggesting that the fading was driven by changes in the accretion rate. We note that it is also possible that a similar evolution in the Comptonisation component took place in the aforementioned cases of IGR~J17494$-$3030 and XTE~J1709$-$267. However, this could not be properly characterised in those studies due to the limited energy coverage of the \xmm\ observations, which made the hard X-ray spectral shape more difficult to constrain.

\subsubsection{\slxtres}\label{subsec:Discburst}

The system exhibited two Type-I X-ray bursts. Both events had durations of about $30$~s and displayed clear spectral softening, with the BB temperature decreasing from $\sim 2$~keV to $\sim 0.7$~keV during the decay, as typically observed during the cooling phase of the NS photosphere following thermonuclear ignition \citep[see e.g.][]{Lewin1993SSRv...62..223L}. The burst fluences were of the order of $f_b = 5-7 \times 10^{-8}~\fb$. We found no evidence of photospheric radius expansion in either burst, consistent with previous observations \citep{Chelovekov2017AstL}.

The detection of two bursts establishes an upper limit on the recurrence time of $\tau_{\rm rec} \lesssim 5$~h during this epoch, refining the previous constraint of $<7.4$~h reported by \citet{Chelovekov2017AstL}. We note that additional bursts may have gone undetected during data gaps caused by Earth occultations. The estimation of the recurrence time helps constrain the burning regime, and thus the fuel composition. The time typically required to exhaust accreted hydrogen via stable burning is $t_{H} \approx 11\,\mathrm{h}\left(\frac{0.02}{Z_{\mathrm{CNO}}}\right)\left(\frac{X}{0.7}\right)$, where $Z_{\mathrm{CNO}}$ is the CNO mass fraction and $X$ the hydrogen mass fraction. For solar-type metallicities, our derived $\tau_{\rm rec} \lesssim 5$~h is significantly shorter than $t_{H}$, implying that a substantial fraction of hydrogen remains at the onset of the thermonuclear runaway. This favours mixed H/He burning \citep[see, e.g.,][]{Cumming2004NuPhS.132..435C}.%
\begin{figure}[!htbp]
\centering
\includegraphics[width=\columnwidth]{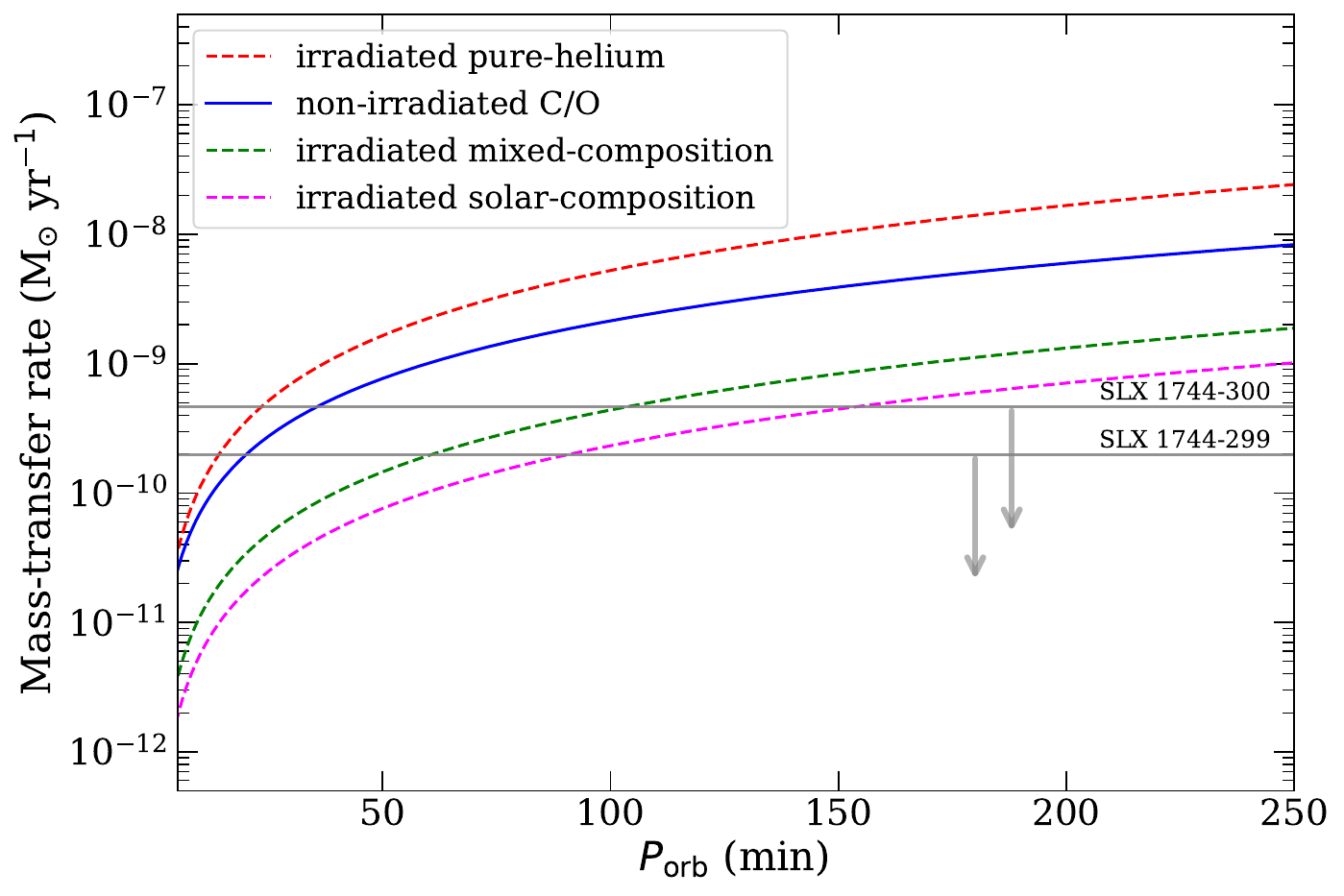}
\caption{Stability limits for non-irradiated C/O discs, irradiated mixed-composition discs and irradiated solar-composition discs based on \citet{Menou2002ApJ...564L..81M} and \citet{Lasota2008A&A...486..523L}. In both cases, the upper limits on the mass transfer rate reported in this work yield upper limits on $P_{\rm orb}$.}
\label{fig:DIM}
\end{figure}%
In this case, we can adopt a representative value of $\alpha \sim 40$, typically observed in mixed H/He bursts \citep{Galloway2008ApJS..179..360G}, to obtain an order-of-magnitude estimate of the persistent bolometric flux. The $\alpha$ parameter is defined as the ratio between the accretion fluence accumulated between bursts and the burst fluence itself \citep[$\alpha = F_{\rm pers}\Delta t / f_{\rm b}$;][]{Lewin1993SSRv...62..223L}. Using the measured bolometric burst fluences together with the observed recurrence time between the two bursts ($\Delta t \simeq 1.82 \times 10^{4}~{\rm s}$), we infer a bolometric persistent flux of $\sim(1.1-1.4)\times10^{-10}~\flux$. Assuming the distance upper limit of $d<10.3~{\rm kpc}$, this corresponds to $L_{\rm pers}\lesssim1.8\times10^{36}~\lum$, in good agreement with the luminosity independently inferred from the spectral fits to the persistent emission.

\subsection{Thermal–viscous stability constraints on the $P_{\mathrm{orb}}$ limits}

One of the few indirect diagnostics available to identify UCXB candidates in the absence of direct $P_{\rm orb}$ measurements is their persistent behaviour at low accretion rates. According to the DIM (\citealt{Lasota2001NewAR..45..449L}; see also \citealt{Coriat2012MNRAS.424.1991C}), systems accreting at very low rates should be transient unless they have very short orbital periods, even within the ultracompact regime. This would allow their discs to remain fully ionised (i.e. thermally stable) even at low $\dot{M}$ \citep{intZand2007A&A...465..953I}.
To investigate whether this could be the case for \slxdos\ and \slxtres, we estimated their mass accretion rates ($\dot{M}_{\rm acc}$) from their observed persistent X-ray luminosities using $L_{\rm X} = \eta c^{2} \dot{M}_{\rm acc}$ and assuming a radiatively efficient scenario (i.e. $\eta = 0.1$; e.g.~\citealt{Coriat2012MNRAS.424.1991C}). These values were then compared with the critical stability thresholds predicted by DIM, in which thermal–viscous stability is governed by the mass-transfer rate from the donor star (assuming $\dot{M}_{\rm transfer} \simeq \dot{M}_{\rm acc}$) and the orbital period ($P_{\rm orb}$).

For \slxdos, the persistently faint X-ray luminosity of $L_{\rm X} \lesssim 1.06\times10^{36}~\lum$ implies $\dot{M}_{\rm acc} \lesssim 1.9\times10^{-10} M_{\odot}~\mathrm{yr^{-1}}$. Comparing the inferred $\dot{M}_{\rm acc}$ with the critical stability thresholds predicted by the DIM, we find that a persistently stable disc at this accretion rate requires an orbital period $P_{\rm orb} \lesssim 90$~min, indicating that an ultracompact nature is a highly plausible scenario (see Fig.~\ref{fig:DIM}). In fact, this low accretion rate, together with the detection of slow-recurrence, intermediate-duration bursts, had already led to the source being proposed as a UCXB candidate and included in catalogues of such systems \citep{intZand2007A&A...465..953I,ArmasPadilla2023A&A...677A.186A}. 

For \slxtres, we inferred an X-ray luminosity of $L_{\rm X} \lesssim 2.56 \times 10^{36}~\lum$, corresponding to $\dot{M}_{\rm acc} \lesssim 4.5\times10^{-10}~M_{\odot}~\mathrm{yr^{-1}}$ and $P_{\rm orb} \lesssim 155$~min.
This constraint on $P_{\rm orb}$ is therefore compatible with a short-period LMXB, including a UCXB. As a matter of fact, if the actual distance were $\lesssim 6.5$~kpc, the inferred mass accretion rate would lie just above the instability threshold for such systems (i.e. $P_{\rm orb} < 80$~min). However, given that $P_{\rm orb}$ values above this threshold are allowed, and considering the properties of the short burst (Sec.~\ref{subsec:Discburst}) detected in this source, we note that \slxtres\ is a significantly weaker UCXB candidate than \slxdos.

The UCXB interpretation can also be compared with the homogeneous \nustar\ study of confirmed UCXBs by \citet{Borghese2026}. In that work, hard-state UCXBs, with luminosities in the $\sim10^{35}-10^{37}~\lum$ range, were generally described with two- or three-component models including one or two thermal components associated with the NS surface or boundary layer, the accretion disc, or both, together with a dominant Comptonised continuum. These systems displayed BB temperatures of $kT_{\rm BB}\sim0.5-1.5$ keV, disc temperatures of $kT_{\rm disc}\sim0.4-0.8$ keV, electron temperatures of $kT_{\rm e}\sim10-40$ keV, and photon indices of $\Gamma\sim1.8-2.0$.  Both \slxdos\ and \slxtres\ broadly fall within the above ranges, although their photon indices ($\Gamma\sim2.3$) are somewhat softer than the average values found in that sample. In particular, the coronal temperature derived for \slxdos\ ($kT_{\rm e}\sim38$ keV) lies well within the range observed in hard-state UCXBs, while \slxtres\ ($kT_{\rm e}\sim9$ keV) lies at the lower end of the distribution.

A notable difference, however, is that our \nustar\ spectra do not statistically require additional thermal components, and any such contribution remains minimal when included. In the homogeneous UCXB study by \citet{Borghese2026}, only two sources, IGR J17494--3030 and 47 Tuc X--9, did not require thermal components. This may be related to their significantly lower luminosities of $\sim10^{34}-10^{35}~\lum$, although poorer counting statistics may also play a role. In this context, the absence of a thermal component in our spectra could also indicate luminosities closer to the $\sim10^{35}~\lum$ regime, which would be possible if the distances to \slxdos\ and \slxtres\ are significantly below the current upper limits. Overall, the comparison remains compatible with a UCXB interpretation for both systems, although the case appears stronger for \slxdos.

\section{Conclusion}

We analysed a \nustar\ observation, which enabled us to separate the contributions of the NS-LMXBs \slxdos\ and \slxtres\ (located only $\sim$2.7~arcmin apart) in the hard energy band for the first time. We find \slxtres\ to be slightly brighter, with a flux ratio of $\sim1.15$ in the 3--78 keV band ($\sim1.3$ when extrapolated to the 0.5--10 keV band). The average spectra of both sources are well described by a thermally Comptonised model, which returned spectral parameters consistent with those of NS-LMXBs during the low-luminosity hard state. This is in agreement with the high fractional rms values ($\sim$21–23 per cent) measured in both cases. The evolution of both sources throughout the $\sim 80$~ks covered by the observation was markedly different. While \slxdos\ showed a constant decrease in flux, likely due to variations in the mass-accretion rate, \slxtres\ remained steady but exhibited two short Type~I bursts consistent with mixed hydrogen/helium burning.

These observations, combined with previously reported upper limits on the distance, yield low persistent X-ray luminosities of $L_{\rm X}\lesssim1.1\times10^{36}~\lum$ and $L_{\rm X}\lesssim2.6\times10^{36}~\lum$ for \slxdos\ and \slxtres, respectively. By comparing the inferred mass-accretion rates (from these luminosities) with the critical values for disc stability predicted by the DIM, we found $P_{\rm orb}\lesssim90$~min and $P_{\rm orb}\lesssim 105$--155~min. While both constraints are technically compatible with the UCXB regime, the former case (i.e. \slxdos) is significantly more robust, not only based on these values but also in light of the intermediate-duration burst previously reported for this source.

\begin{acknowledgements}
We thank the anonymous referee for their careful reading of the manuscript and for their comments, which helped improve the clarity of the paper.
We acknowledge support by Spanish Agencia estatal de investigación via PID2021-124879NB-I00 and PID2024-161863NB-I00.
M.A.P. acknowledges support through the Ramón y Cajal grant RYC2022-035388-I, funded by MCIU/AEI/10.13039/501100011033 and FSE+.
\end{acknowledgements}

\bibliographystyle{aa}
\bibliography{biblio}

@ARTICLE{CruzSanchez2026arXiv260310331C,
       author = {{Cruz-Sanchez}, Nelson and {Saavedra}, Enzo A. and {Fogantini}, Federico A. and {Garc{\'\i}a}, Federico and {Combi}, Jorge A. and {Bachetti}, Matteo and {Imbrogno}, Matteo and {Sidoli}, Lara and {Marino}, Alessio},
        title = "{A candidate proton cyclotron feature in the ultraluminous X-ray source NGC 4656 ULX-1}",
      journal = {\aap},
         year = 2026,
        month = mar,
       volume = {707},
          eid = {L20},
        pages = {L20},
          doi = {10.1051/0004-6361/202658875},
archivePrefix = {arXiv},
       eprint = {2603.10331},
 primaryClass = {astro-ph.HE},
       adsurl = {https://ui.adsabs.harvard.edu/abs/2026A&A...707L..20C}
}

@ARTICLE{Pavlinsky2021A&A...650A..42P,
       author = {{Pavlinsky}, M. and {Tkachenko}, A. and {Levin}, V. and {Alexandrovich}, N. and {Arefiev}, V. and {Babyshkin}, V. and {Batanov}, O. and {Bodnar}, Yu. and {Bogomolov}, A. and {Bubnov}, A. and {Buntov}, M. and {Burenin}, R. and {Chelovekov}, I. and {Chen}, C.-T. and {Drozdova}, T. and {Ehlert}, S. and {Filippova}, E. and {Frolov}, S. and {Gamkov}, D. and {Garanin}, S. and {Garin}, M. and {Glushenko}, A. and {Gorelov}, A. and {Grebenev}, S. and {Grigorovich}, S. and {Gureev}, P. and {Gurova}, E. and {Ilkaev}, R. and {Katasonov}, I. and {Krivchenko}, A. and {Krivonos}, R. and {Korotkov}, F. and {Kudelin}, M. and {Kuznetsova}, M. and {Lazarchuk}, V. and {Lomakin}, I. and {Lapshov}, I. and {Lipilin}, V. and {Lutovinov}, A. and {Mereminskiy}, I. and {Molkov}, S. and {Nazarov}, V. and {Oleinikov}, V. and {Pikalov}, E. and {Ramsey}, B.~D. and {Roiz}, I. and {Rotin}, A. and {Ryadov}, A. and {Sankin}, E. and {Sazonov}, S. and {Sedov}, D. and {Semena}, A. and {Semena}, N. and {Serbinov}, D. and {Shirshakov}, A. and {Shtykovsky}, A. and {Shvetsov}, A. and {Sunyaev}, R. and {Swartz}, D.~A. and {Tambov}, V. and {Voron}, V. and {Yaskovich}, A.},
        title = "{The ART-XC telescope on board the SRG observatory}",
      journal = {\aap},
         year = 2021,
        month = jun,
       volume = {650},
          eid = {A42},
        pages = {A42},
          doi = {10.1051/0004-6361/202040265},
archivePrefix = {arXiv},
       eprint = {2103.12479},
 primaryClass = {astro-ph.HE},
       adsurl = {https://ui.adsabs.harvard.edu/abs/2021A&A...650A..42P}
}

@INPROCEEDINGS{Koglin2011SPIE.8147E..0JK,
       author = {{Koglin}, Jason E. and {An}, HongJun and {Barri{\`e}re}, Nicolas and {Brejnholt}, Nicolai F. and {Christensen}, Finn E. and {Craig}, William W. and {Hailey}, Charles J. and {Jakobsen}, Anders Clemen and {Madsen}, Kristin K. and {Mori}, Kaya and {Nynka}, Melania and {Fernandez-Perea}, Monica and {Pivovaroff}, Michael J. and {Ptak}, Andrew and {Sleator}, Clio and {Thornhill}, Doug and {Vogel}, Julia K. and {Wik}, Daniel R. and {Zhang}, William W.},
        title = "{First results from the ground calibration of the NuSTAR flight optics}",
    booktitle = {Society of Photo-Optical Instrumentation Engineers (SPIE) Conference Series},
         year = 2011,
       editor = {{O'Dell}, Stephen L. and {Pareschi}, Giovanni},
       series = {Society of Photo-Optical Instrumentation Engineers (SPIE) Conference Series},
       volume = {8147},
        month = sep,
          eid = {81470J},
        pages = {81470J},
          doi = {10.1117/12.895279},
       adsurl = {https://ui.adsabs.harvard.edu/abs/2011SPIE.8147E..0JK}
}

@INPROCEEDINGS{Blackburn1995ASPC...77..367B,
       author = {{Blackburn}, J.~K.},
        title = "{FTOOLS: A FITS Data Processing and Analysis Software Package}",
    booktitle = {Astronomical Data Analysis Software and Systems IV},
         year = 1995,
       editor = {{Shaw}, R.~A. and {Payne}, H.~E. and {Hayes}, J.~J.~E.},
       series = {Astronomical Society of the Pacific Conference Series},
       volume = {77},
        month = jan,
        pages = {367},
       adsurl = {https://ui.adsabs.harvard.edu/abs/1995ASPC...77..367B}
}

@ARTICLE{CruzSanchez2026A&A...705A.136C,
       author = {{Cruz-Sanchez}, N. and {Saavedra}, E.~A. and {Fogantini}, F.~A. and {Garc{\'\i}a}, F. and {Combi}, J.~A.},
        title = "{The hard ultraluminous state of NGC 5055 ULX X-1}",
      journal = {\aap},
         year = 2026,
        month = jan,
       volume = {705},
          eid = {A136},
        pages = {A136},
          doi = {10.1051/0004-6361/202556922},
archivePrefix = {arXiv},
       eprint = {2511.13686},
 primaryClass = {astro-ph.HE},
       adsurl = {https://ui.adsabs.harvard.edu/abs/2026A&A...705A.136C}
}

@ARTICLE{Done2007A&ARv..15....1D,
       author = {{Done}, Chris and {Gierli{\'n}ski}, Marek and {Kubota}, Aya},
        title = "{Modelling the behaviour of accretion flows in X-ray binaries. Everything you always wanted to know about accretion but were afraid to ask}",
      journal = {\aapr},
         year = 2007,
        month = dec,
       volume = {15},
       number = {1},
        pages = {1-66},
          doi = {10.1007/s00159-007-0006-1},
archivePrefix = {arXiv},
       eprint = {0708.0148},
 primaryClass = {astro-ph},
       adsurl = {https://ui.adsabs.harvard.edu/abs/2007A&ARv..15....1D}
}

@ARTICLE{Matranga2017A&A...603A..39M,
       author = {{Matranga}, M. and {Papitto}, A. and {Di Salvo}, T. and {Bozzo}, E. and {Torres}, D.~F. and {Iaria}, R. and {Burderi}, L. and {Rea}, N. and {de Martino}, D. and {Sanchez-Fernandez}, C. and {Gambino}, A.~F. and {Ferrigno}, C. and {Stella}, L.},
        title = "{XMM-Newton and INTEGRAL view of the hard state of EXO 1745-248 during its 2015 outburst}",
      journal = {\aap},
         year = 2017,
        month = jul,
       volume = {603},
          eid = {A39},
        pages = {A39},
          doi = {10.1051/0004-6361/201629731},
archivePrefix = {arXiv},
       eprint = {1703.07389},
 primaryClass = {astro-ph.HE},
       adsurl = {https://ui.adsabs.harvard.edu/abs/2017A&A...603A..39M}
}

@ARTICLE{vandenEijnden2018MNRAS.475.2027V,
       author = {{van den Eijnden}, J. and {Degenaar}, N. and {Pinto}, C. and {Patruno}, A. and {Wette}, K. and {Messenger}, C. and {Hern{\'a}ndez Santisteban}, J.~V. and {Wijnands}, R. and {Miller}, J.~M. and {Altamirano}, D. and {Paerels}, F. and {Chakrabarty}, D. and {Fabian}, A.~C.},
        title = "{The very faint X-ray binary IGR J17062-6143: a truncated disc, no pulsations, and a possible outflow}",
      journal = {\mnras},
         year = 2018,
        month = apr,
       volume = {475},
       number = {2},
        pages = {2027-2044},
          doi = {10.1093/mnras/stx3224},
       adsurl = {https://ui.adsabs.harvard.edu/abs/2018MNRAS.475.2027V}
}

@ARTICLE{Degenaar2015MNRAS.451L..85D,
       author = {{Degenaar}, N. and {Miller}, J.~M. and {Chakrabarty}, D. and {Harrison}, F.~A. and {Kara}, E. and {Fabian}, A.~C.},
        title = "{A NuSTAR observation of disc reflection from close to the neutron star in 4U 1608-52.}",
      journal = {\mnras},
         year = 2015,
        month = jul,
       volume = {451},
        pages = {L85-L89},
          doi = {10.1093/mnrasl/slv072},
archivePrefix = {arXiv},
       eprint = {1505.07112},
 primaryClass = {astro-ph.HE},
       adsurl = {https://ui.adsabs.harvard.edu/abs/2015MNRAS.451L..85D}
}

@ARTICLE{Ludlam2017ApJ...836..140L,
       author = {{Ludlam}, R.~M. and {Miller}, J.~M. and {Bachetti}, M. and {Barret}, D. and {Bostrom}, A.~C. and {Cackett}, E.~M. and {Degenaar}, N. and {Di Salvo}, T. and {Natalucci}, L. and {Tomsick}, J.~A. and {Paerels}, F. and {Parker}, M.~L.},
        title = "{A Hard Look at the Neutron Stars and Accretion Disks in 4U 1636-53, GX 17+2, and 4U 1705-44 with NuStar}",
      journal = {\apj},
         year = 2017,
        month = feb,
       volume = {836},
       number = {1},
          eid = {140},
        pages = {140},
          doi = {10.3847/1538-4357/836/1/140},
archivePrefix = {arXiv},
       eprint = {1701.01774},
 primaryClass = {astro-ph.HE},
       adsurl = {https://ui.adsabs.harvard.edu/abs/2017ApJ...836..140L}
}

@ARTICLE{Cumming2004NuPhS.132..435C,
       author = {{Cumming}, Andrew},
        title = "{Thermonuclear X-ray bursts: theory vs. observations}",
      journal = {Nuclear Physics B Proceedings Supplements},
         year = 2004,
        month = jun,
       volume = {132},
        pages = {435-445},
          doi = {10.1016/j.nuclphysbps.2004.04.078},
archivePrefix = {arXiv},
       eprint = {astro-ph/0309626},
 primaryClass = {astro-ph},
       adsurl = {https://ui.adsabs.harvard.edu/abs/2004NuPhS.132..435C}
}

@ARTICLE{Baharamian2017MNRAS.467.2199B,
       author = {{Bahramian}, Arash and {Heinke}, Craig O. and {Tudor}, Vlad and {Miller-Jones}, James C.~A. and {Bogdanov}, Slavko and {Maccarone}, Thomas J. and {Knigge}, Christian and {Sivakoff}, Gregory R. and {Chomiuk}, Laura and {Strader}, Jay and {Garcia}, Javier A. and {Kallman}, Timothy},
        title = "{The ultracompact nature of the black hole candidate X-ray binary 47 Tuc X9}",
      journal = {\mnras},
         year = 2017,
        month = may,
       volume = {467},
       number = {2},
        pages = {2199-2216},
          doi = {10.1093/mnras/stx166},
archivePrefix = {arXiv},
       eprint = {1702.02167},
 primaryClass = {astro-ph.HE},
       adsurl = {https://ui.adsabs.harvard.edu/abs/2017MNRAS.467.2199B}
}

@ARTICLE{makishima1986ApJ...308..635M,
       author = {{Makishima}, K. and {Maejima}, Y. and {Mitsuda}, K. and {Bradt}, H.~V. and {Remillard}, R.~A. and {Tuohy}, I.~R. and {Hoshi}, R. and {Nakagawa}, M.},
        title = "{Simultaneous X-Ray and Optical Observations of GX 339-4 in an X-Ray High State}",
      journal = {\apj},
         year = 1986,
        month = sep,
       volume = {308},
        pages = {635},
          doi = {10.1086/164534},
       adsurl = {https://ui.adsabs.harvard.edu/abs/1986ApJ...308..635M}
}

@ARTICLE{mitsuda1984PASJ...36..741M,
       author = {{Mitsuda}, K. and {Inoue}, H. and {Koyama}, K. and {Makishima}, K. and {Matsuoka}, M. and {Ogawara}, Y. and {Shibazaki}, N. and {Suzuki}, K. and {Tanaka}, Y. and {Hirano}, T.},
        title = "{Energy spectra of low-mass binary X-ray sources observed from Tenma.}",
      journal = {\pasj},
         year = 1984,
        month = jan,
       volume = {36},
        pages = {741-759},
       adsurl = {https://ui.adsabs.harvard.edu/abs/1984PASJ...36..741M}
}

@ARTICLE{zdziarski1996MNRAS.283..193Z,
       author = {{Zdziarski}, A.~A. and {Johnson}, W.~N. and {Magdziarz}, P.},
        title = "{Broad-band {\ensuremath{\gamma}}-ray and X-ray spectra of NGC 4151 and their implications for physical processes and geometry.}",
      journal = {\mnras},
         year = 1996,
        month = nov,
       volume = {283},
       number = {1},
        pages = {193-206},
          doi = {10.1093/mnras/283.1.193},
archivePrefix = {arXiv},
       eprint = {astro-ph/9607015},
 primaryClass = {astro-ph},
       adsurl = {https://ui.adsabs.harvard.edu/abs/1996MNRAS.283..193Z}
}

@ARTICLE{zycki1999MNRAS.309..561Z,
       author = {{{\.Z}ycki}, Piotr T. and {Done}, Chris and {Smith}, David A.},
        title = "{The 1989 May outburst of the soft X-ray transient GS 2023+338 (V404 Cyg)}",
      journal = {\mnras},
         year = 1999,
        month = nov,
       volume = {309},
       number = {3},
        pages = {561-575},
          doi = {10.1046/j.1365-8711.1999.02885.x},
archivePrefix = {arXiv},
       eprint = {astro-ph/9904304},
 primaryClass = {astro-ph},
       adsurl = {https://ui.adsabs.harvard.edu/abs/1999MNRAS.309..561Z}
}

@ARTICLE{MunozDarias2011MNRAS.410..679M,
       author = {{Mu{\~n}oz-Darias}, T. and {Motta}, S. and {Belloni}, T.~M.},
        title = "{Fast variability as a tracer of accretion regimes in black hole transients}",
      journal = {\mnras},
         year = 2011,
        month = jan,
       volume = {410},
       number = {1},
        pages = {679-684},
          doi = {10.1111/j.1365-2966.2010.17476.x},
archivePrefix = {arXiv},
       eprint = {1008.0558},
 primaryClass = {astro-ph.HE},
       adsurl = {https://ui.adsabs.harvard.edu/abs/2011MNRAS.410..679M}
}

@ARTICLE{Borghese2026,
  author  = {{Borghese}, A. and {Armas Padilla}, M. and {Muñoz-Darias}, T.},
  title   = {The NuSTAR view of ultra-compact X-ray binaries},
  journal = {\aap},
  volume  = {710},
  pages   = {A43},
  year    = {2026}
}

@ARTICLE{Bachetti2018ApJ...853L..21B,
       author = {{Bachetti}, Matteo and {Huppenkothen}, Daniela},
        title = "{No Time for Dead Time: Use the Fourier Amplitude Differences to Normalize Dead-time-affected Periodograms}",
      journal = {\apjl},
         year = 2018,
        month = feb,
       volume = {853},
       number = {2},
          eid = {L21},
        pages = {L21},
          doi = {10.3847/2041-8213/aaa83b},
archivePrefix = {arXiv},
       eprint = {1709.09700},
 primaryClass = {astro-ph.HE},
       adsurl = {https://ui.adsabs.harvard.edu/abs/2018ApJ...853L..21B}
}

@ARTICLE{MunozDarias2014MNRAS.443.3270M,
       author = {{Mu{\~n}oz-Darias}, T. and {Fender}, R.~P. and {Motta}, S.~E. and {Belloni}, T.~M.},
        title = "{Black hole-like hysteresis and accretion states in neutron star low-mass X-ray binaries}",
      journal = {\mnras},
         year = 2014,
        month = oct,
       volume = {443},
       number = {4},
        pages = {3270-3283},
          doi = {10.1093/mnras/stu1334},
archivePrefix = {arXiv},
       eprint = {1407.1318},
 primaryClass = {astro-ph.HE},
       adsurl = {https://ui.adsabs.harvard.edu/abs/2014MNRAS.443.3270M}
}

@ARTICLE{Coriat2012MNRAS.424.1991C,
       author = {{Coriat}, M. and {Fender}, R.~P. and {Dubus}, G.},
        title = "{Revisiting a fundamental test of the disc instability model for X-ray binaries}",
      journal = {\mnras},
         year = 2012,
        month = aug,
       volume = {424},
       number = {3},
        pages = {1991-2001},
          doi = {10.1111/j.1365-2966.2012.21339.x},
archivePrefix = {arXiv},
       eprint = {1205.5038},
 primaryClass = {astro-ph.HE},
       adsurl = {https://ui.adsabs.harvard.edu/abs/2012MNRAS.424.1991C}
}

@ARTICLE{Chelovekov2017AstL,
       author = {{Chelovekov}, I.~V. and {Grebenev}, S.~A. and {Mereminskiy}, I.~A. and {Prosvetov}, A.~V.},
        title = "{Type I X-ray Bursts Detected by the JEM-X Telescope Onboard the INTEGRAL Observatory in 2003-2015}",
      journal = {Astronomy Letters},
         year = 2017,
        month = dec,
       volume = {43},
       number = {12},
        pages = {781-795},
          doi = {10.1134/S1063773717120076},
       adsurl = {https://ui.adsabs.harvard.edu/abs/2017AstL...43..781C}
}

@ARTICLE{Madsen2020arXiv200500569M,
       author = {{Madsen}, Kristin K. and {Grefenstette}, Brian W. and {Pike}, Sean and {Miyasaka}, Hiromasa and {Brightman}, Murray and {Forster}, Karl and {Harrison}, Fiona A.},
        title = "{NuSTAR low energy effective area correction due to thermal blanket tear}",
      journal = {arXiv e-prints},
         year = 2020,
        month = may,
          eid = {arXiv:2005.00569},
        pages = {arXiv:2005.00569},
          doi = {10.48550/arXiv.2005.00569},
archivePrefix = {arXiv},
       eprint = {2005.00569},
 primaryClass = {astro-ph.IM},
       adsurl = {https://ui.adsabs.harvard.edu/abs/2020arXiv200500569M}
}

@ARTICLE{Madsen2022JATIS...8c4003M,
       author = {{Madsen}, Kristin K. and {Forster}, Karl and {Grefenstette}, Brian and {Harrison}, Fiona A. and {Miyasaka}, Hiromasa},
        title = "{Effective area calibration of the Nuclear Spectroscopic Telescope Array}",
      journal = {Journal of Astronomical Telescopes, Instruments, and Systems},
         year = 2022,
        month = jul,
       volume = {8},
          eid = {034003},
        pages = {034003},
          doi = {10.1117/1.JATIS.8.3.034003},
archivePrefix = {arXiv},
       eprint = {2110.11522},
 primaryClass = {astro-ph.IM},
       adsurl = {https://ui.adsabs.harvard.edu/abs/2022JATIS...8c4003M}
}

@ARTICLE{Madsen2015ApJS..220....8M,
       author = {{Madsen}, Kristin K. and {Harrison}, Fiona A. and {Markwardt}, Craig B. and {An}, Hongjun and {Grefenstette}, Brian W. and {Bachetti}, Matteo and {Miyasaka}, Hiromasa and {Kitaguchi}, Takao and {Bhalerao}, Varun and {Boggs}, Steve and {Christensen}, Finn E. and {Craig}, William W. and {Forster}, Karl and {Fuerst}, Felix and {Hailey}, Charles J. and {Perri}, Matteo and {Puccetti}, Simonetta and {Rana}, Vikram and {Stern}, Daniel and {Walton}, Dominic J. and {J{\o}rgen Westergaard}, Niels and {Zhang}, William W.},
        title = "{Calibration of the NuSTAR High-energy Focusing X-ray Telescope.}",
      journal = {\apjs},
         year = 2015,
        month = sep,
       volume = {220},
       number = {1},
          eid = {8},
        pages = {8},
          doi = {10.1088/0067-0049/220/1/8},
archivePrefix = {arXiv},
       eprint = {1504.01672},
 primaryClass = {astro-ph.IM},
       adsurl = {https://ui.adsabs.harvard.edu/abs/2015ApJS..220....8M}
}

@ARTICLE{Diez2023A&A...674A.147D,
       author = {{Diez}, C.~M. and {Grinberg}, V. and {F{\"u}rst}, F. and {El Mellah}, I. and {Zhou}, M. and {Santangelo}, A. and {Mart{\'\i}nez-N{\'u}{\~n}ez}, S. and {Amato}, R. and {Hell}, N. and {Kretschmar}, P.},
        title = "{Observing the onset of the accretion wake in Vela X-1}",
      journal = {\aap},
         year = 2023,
        month = jun,
       volume = {674},
          eid = {A147},
        pages = {A147},
          doi = {10.1051/0004-6361/202245708},
archivePrefix = {arXiv},
       eprint = {2303.09631},
 primaryClass = {astro-ph.HE},
       adsurl = {https://ui.adsabs.harvard.edu/abs/2023A&A...674A.147D}
}

@ARTICLE{Zalot2024A&A...686A..95Z,
       author = {{Zalot}, Nicolas and {Sokolova-Lapa}, Ekaterina and {Stierhof}, Jakob and {Ballhausen}, Ralf and {Zainab}, Aafia and {Pottschmidt}, Katja and {F{\"u}rst}, Felix and {Thalhammer}, Philipp and {Islam}, Nazma and {Diez}, Camille M. and {Kretschmar}, Peter and {Berger}, Katrin and {Rothschild}, Richard and {Malacaria}, Christian and {Pradhan}, Pragati and {Wilms}, J{\"o}rn},
        title = "{An in-depth analysis of the variable cyclotron lines in GX 301{\ensuremath{-}}2}",
      journal = {\aap},
         year = 2024,
        month = jun,
       volume = {686},
          eid = {A95},
        pages = {A95},
          doi = {10.1051/0004-6361/202348841},
archivePrefix = {arXiv},
       eprint = {2403.11682},
 primaryClass = {astro-ph.HE},
       adsurl = {https://ui.adsabs.harvard.edu/abs/2024A&A...686A..95Z}
}

@ARTICLE{Ballhausen2024A&A...688A.214B,
       author = {{Ballhausen}, R. and {Thalhammer}, P. and {Pradhan}, P. and {Sokolova-Lapa}, E. and {Stierhof}, J. and {Pottschmidt}, K. and {Wilms}, J. and {Coley}, J.~B. and {Kretschmar}, P. and {F{\"u}rst}, F. and {Becker}, P. and {West}, B. and {Malacaria}, C. and {Wolff}, M.~T. and {Rothschild}, R. and {Staubert}, R.},
        title = "{The giant outburst of EXO 2030+375. II. Broadband spectroscopy and evolution}",
      journal = {\aap},
         year = 2024,
        month = aug,
       volume = {688},
          eid = {A214},
        pages = {A214},
          doi = {10.1051/0004-6361/202348595},
archivePrefix = {arXiv},
       eprint = {2406.13029},
 primaryClass = {astro-ph.HE},
       adsurl = {https://ui.adsabs.harvard.edu/abs/2024A&A...688A.214B}
}

@ARTICLE{LaMonca2024ApJ...960L..11L,
       author = {{La Monaca}, Fabio and {Di Marco}, Alessandro and {Poutanen}, Juri and {Bachetti}, Matteo and {Motta}, Sara Elisa and {Papitto}, Alessandro and {Pilia}, Maura and {Xie}, Fei and {Bianchi}, Stefano and {Bobrikova}, Anna and {Costa}, Enrico and {Deng}, Wei and {Ge}, Ming-Yu and {Illiano}, Giulia and {Jia}, Shu-Mei and {Krawczynski}, Henric and {Lai}, Eleonora Veronica and {Liu}, Kuan and {Mastroserio}, Guglielmo and {Muleri}, Fabio and {Rankin}, John and {Soffitta}, Paolo and {Veledina}, Alexandra and {Ambrosino}, Filippo and {Del Santo}, Melania and {Chen}, Wei and {Garcia}, Javier A. and {Kaaret}, Philip and {Russell}, Thomas D. and {Wei}, Wen-Hao and {Zhang}, Shuang-Nan and {Zuo}, Chao and {Arzoumanian}, Zaven and {Cocchi}, Massimo and {Gnarini}, Andrea and {Farinelli}, Ruben and {Gendreau}, Keith and {Ursini}, Francesco and {Weisskopf}, Martin C. and {Zane}, Silvia and {Agudo}, Iv{\'a}n and {Antonelli}, Lucio A. and {Baldini}, Luca and {Baumgartner}, Wayne H. and {Bellazzini}, Ronaldo and {Bongiorno}, Stephen D. and {Bonino}, Raffaella and {Brez}, Alessandro and {Bucciantini}, Niccol{\`o} and {Capitanio}, Fiamma and {Castellano}, Simone and {Cavazzuti}, Elisabetta and {Chen}, Chien-Ting and {Ciprini}, Stefano and {De Rosa}, Alessandra and {Del Monte}, Ettore and {Di Gesu}, Laura and {Di Lalla}, Niccol{\`o} and {Donnarumma}, Immacolata and {Doroshenko}, Victor and {Dov{\v{c}}iak}, Michal and {Ehlert}, Steven R. and {Enoto}, Teruaki and {Evangelista}, Yuri and {Fabiani}, Sergio and {Ferrazzoli}, Riccardo and {Gunji}, Shuichi and {Hayashida}, Kiyoshi and {Heyl}, Jeremy and {Iwakiri}, Wataru and {Jorstad}, Svetlana G. and {Karas}, Vladimir and {Kislat}, Fabian and {Kitaguchi}, Takao and {Kolodziejczak}, Jeffery J. and {Latronico}, Luca and {Liodakis}, Ioannis and {Maldera}, Simone and {Manfreda}, Alberto and {Marin}, Fr{\'e}d{\'e}ric and {Marinucci}, Andrea and {Marscher}, Alan P. and {Marshall}, Herman L. and {Massaro}, Francesco and {Matt}, Giorgio and {Mitsuishi}, Ikuyuki and {Mizuno}, Tsunefumi and {Negro}, Michela and {Ng}, Chi-Yung and {O'Dell}, Stephen L. and {Omodei}, Nicola and {Oppedisano}, Chiara and {Pavlov}, George G. and {Peirson}, Abel L. and {Perri}, Matteo and {Pesce-Rollins}, Melissa and {Petrucci}, Pierre-Olivier and {Possenti}, Andrea and {Puccetti}, Simonetta and {Ramsey}, Brian D. and {Ratheesh}, Ajay and {Roberts}, Oliver J. and {Romani}, Roger W. and {Sgr{\`o}}, Carmelo and {Slane}, Patrick and {Spandre}, Gloria and {Swartz}, Douglas A. and {Tamagawa}, Toru and {Tavecchio}, Fabrizio and {Taverna}, Roberto and {Tawara}, Yuzuru and {Tennant}, Allyn F. and {Thomas}, Nicholas E. and {Tombesi}, Francesco and {Trois}, Alessio and {Tsygankov}, Sergey S. and {Turolla}, Roberto and {Vink}, Jacco and {Wu}, Kinwah and {IXPE Collaboration}},
        title = "{Highly Significant Detection of X-Ray Polarization from the Brightest Accreting Neutron Star Sco X-1}",
      journal = {\apjl},
         year = 2024,
        month = jan,
       volume = {960},
       number = {2},
          eid = {L11},
        pages = {L11},
          doi = {10.3847/2041-8213/ad132d},
archivePrefix = {arXiv},
       eprint = {2311.06359},
 primaryClass = {astro-ph.HE},
       adsurl = {https://ui.adsabs.harvard.edu/abs/2024ApJ...960L..11L}
}

@ARTICLE{Grefenstette2022arXiv220604058G,
       author = {{Grefenstette}, Brian and {Brightman}, Murray and {Earnshaw}, Hannah P. and {Forster}, Karl and {Madsen}, Kristin K. and {Miyasaka}, Hiromasa},
        title = "{Measuring the Evolution of the NuSTAR Detector Gains}",
      journal = {arXiv e-prints},
         year = 2022,
        month = jun,
          eid = {arXiv:2206.04058},
        pages = {arXiv:2206.04058},
          doi = {10.48550/arXiv.2206.04058},
archivePrefix = {arXiv},
       eprint = {2206.04058},
 primaryClass = {astro-ph.IM},
       adsurl = {https://ui.adsabs.harvard.edu/abs/2022arXiv220604058G}
}

@ARTICLE{Goodman2010CAMCS...5...65G,
       author = {{Goodman}, Jonathan and {Weare}, Jonathan},
        title = "{Ensemble samplers with affine invariance}",
      journal = {Communications in Applied Mathematics and Computational Science},
         year = 2010,
        month = jan,
       volume = {5},
       number = {1},
        pages = {65-80},
          doi = {10.2140/camcos.2010.5.65},
       adsurl = {https://ui.adsabs.harvard.edu/abs/2010CAMCS...5...65G}
}

@ARTICLE{intZand2007A&A...465..953I,
       author = {{in't Zand}, J.~J.~M. and {Jonker}, P.~G. and {Markwardt}, C.~B.},
        title = "{Six new candidate ultracompact X-ray binaries}",
      journal = {\aap},
         year = 2007,
        month = apr,
       volume = {465},
       number = {3},
        pages = {953-963},
          doi = {10.1051/0004-6361:20066678},
archivePrefix = {arXiv},
       eprint = {astro-ph/0701810},
 primaryClass = {astro-ph},
       adsurl = {https://ui.adsabs.harvard.edu/abs/2007A&A...465..953I}
}

@ARTICLE{Skinner1990MNRAS.243...72S,
       author = {{Skinner}, G.~K. and {Foster}, A.~J. and {Willmore}, A.~P. and {Eyles}, C.~J.},
        title = "{Localization of one of the galactic centre X-ray burst sources.}",
      journal = {\mnras},
         year = 1990,
        month = mar,
       volume = {243},
        pages = {72-77},
          doi = {10.1093/mnras/243.1.72},
       adsurl = {https://ui.adsabs.harvard.edu/abs/1990MNRAS.243...72S}
}

@ARTICLE{Pavlinsky1994ApJ,
       author = {{Pavlinsky}, M.~N. and {Grebenev}, S.~A. and {Sunyaev}, R.~A.},
        title = "{X-Ray Images of the Galactic Center Obtained with Art-P/Granat: Discovery of New Sources, Variability of Persistent Sources, and Localization of X-Ray Bursters}",
      journal = {\apj},
         year = 1994,
        month = apr,
       volume = {425},
        pages = {110},
          doi = {10.1086/173967},
       adsurl = {https://ui.adsabs.harvard.edu/abs/1994ApJ...425..110P}
}

@ARTICLE{Fortin2024,
  author  = {{Fortin}, F. and {Kalsi}, A. and {García}, F. and {Simaz-Bunzel}, A. and {Chaty}, S.},
  title   = {A catalogue of low-mass X-ray binaries in the Galaxy: from the INTEGRAL to the Gaia era},
  journal = {\aap},
  volume  = {684},
  pages   = {A124},
  year    = {2024}
}

@ARTICLE{Galloway2008ApJS..179..360G,
       author = {{Galloway}, Duncan K. and {Muno}, Michael P. and {Hartman}, Jacob M. and {Psaltis}, Dimitrios and {Chakrabarty}, Deepto},
        title = "{Thermonuclear (Type I) X-Ray Bursts Observed by the Rossi X-Ray Timing Explorer}",
      journal = {\apjs},
         year = 2008,
        month = dec,
       volume = {179},
       number = {2},
        pages = {360-422},
          doi = {10.1086/592044},
archivePrefix = {arXiv},
       eprint = {astro-ph/0608259},
 primaryClass = {astro-ph},
       adsurl = {https://ui.adsabs.harvard.edu/abs/2008ApJS..179..360G}
}

@ARTICLE{Lewin1993SSRv...62..223L,
       author = {{Lewin}, Walter H.~G. and {van Paradijs}, Jan and {Taam}, Ronald E.},
        title = "{X-Ray Bursts}",
      journal = {\ssr},
         year = 1993,
        month = sep,
       volume = {62},
       number = {3-4},
        pages = {223-389},
          doi = {10.1007/BF00196124},
       adsurl = {https://ui.adsabs.harvard.edu/abs/1993SSRv...62..223L}
}

@ARTICLE{intZand2011A&A...525A.111I,
       author = {{in't Zand}, J.~J.~M. and {Galloway}, D.~K. and {Ballantyne}, D.~R.},
        title = "{Achromatic late-time variability in thermonuclear X-ray bursts. An accretion disk disrupted by a nova-like shell?}",
      journal = {\aap},
         year = 2011,
        month = jan,
       volume = {525},
          eid = {A111},
        pages = {A111},
          doi = {10.1051/0004-6361/201015556},
archivePrefix = {arXiv},
       eprint = {1009.5359},
 primaryClass = {astro-ph.HE},
       adsurl = {https://ui.adsabs.harvard.edu/abs/2011A&A...525A.111I}
}

@ARTICLE{Degenaar2016MNRAS.456.4256D,
       author = {{Degenaar}, N. and {Koljonen}, K.~I.~I. and {Chakrabarty}, D. and {Kara}, E. and {Altamirano}, D. and {Miller}, J.~M. and {Fabian}, A.~C.},
        title = "{Probing the effects of a thermonuclear X-ray burst on the neutron star accretion flow with NuSTAR}",
      journal = {\mnras},
         year = 2016,
        month = mar,
       volume = {456},
       number = {4},
        pages = {4256-4265},
          doi = {10.1093/mnras/stv2965},
archivePrefix = {arXiv},
       eprint = {1601.01680},
 primaryClass = {astro-ph.HE},
       adsurl = {https://ui.adsabs.harvard.edu/abs/2016MNRAS.456.4256D}
}

@MISC{hendrics,
       author = {{Bachetti}, Matteo},
        title = "{HENDRICS: High ENergy Data Reduction Interface from the Command Shell}",
 howpublished = {Astrophysics Source Code Library, record ascl:1805.019},
         year = 2018,
        month = may,
          eid = {ascl:1805.019},
        pages = {ascl:1805.019},
archivePrefix = {ascl},
       eprint = {1805.019},
       adsurl = {https://ui.adsabs.harvard.edu/abs/2018ascl.soft05019B}
}

@ARTICLE{stingray2019ApJ...881...39H,
       author = {{Huppenkothen}, Daniela and {Bachetti}, Matteo and {Stevens}, Abigail L. and {Migliari}, Simone and {Balm}, Paul and {Hammad}, Omar and {Khan}, Usman Mahmood and {Mishra}, Himanshu and {Rashid}, Haroon and {Sharma}, Swapnil and {Martinez Ribeiro}, Evandro and {Valles Blanco}, Ricardo},
        title = "{Stingray: A Modern Python Library for Spectral Timing}",
      journal = {\apj},
         year = 2019,
        month = aug,
       volume = {881},
       number = {1},
          eid = {39},
        pages = {39},
          doi = {10.3847/1538-4357/ab258d},
archivePrefix = {arXiv},
       eprint = {1901.07681},
 primaryClass = {astro-ph.IM},
       adsurl = {https://ui.adsabs.harvard.edu/abs/2019ApJ...881...39H}
}

@ARTICLE{2000ApJ...542..914W,
       author = {{Wilms}, J. and {Allen}, A. and {McCray}, R.},
        title = "{On the Absorption of X-Rays in the Interstellar Medium}",
      journal = {\apj},
         year = 2000,
        month = oct,
       volume = {542},
       number = {2},
        pages = {914-924},
          doi = {10.10/317016},
archivePrefix = {arXiv},
       eprint = {astro-ph/0008425},
 primaryClass = {astro-ph}
}

@ARTICLE{1996ApJ...465..487V,
       author = {{Verner}, D.~A. and {Ferland}, G.~J. and {Korista}, K.~T. and {Yakovlev}, D.~G.},
        title = "{Atomic Data for Astrophysics. II. New Analytic FITS for Photoionization Cross Sections of Atoms and Ions}",
      journal = {\apj},
         year = 1996,
        month = jul,
       volume = {465},
        pages = {487},
          doi = {10.10/177435},
archivePrefix = {arXiv},
       eprint = {astro-ph/9601009},
 primaryClass = {astro-ph}
}

@ARTICLE{Mori2005AdSpR..35.1137M,
       author = {{Mori}, Hideyuki and {Maeda}, Yoshitomo and {Pavlov}, George G. and {Sakano}, Masaaki and {Tsuboi}, Yohko},
        title = "{XMM-Newton observations of the Mouse, SLX 1744 299 and SLX 1744 300}",
      journal = {Advances in Space Research},
         year = 2005,
        month = jan,
       volume = {35},
       number = {6},
        pages = {1137-1141},
          doi = {10.1016/j.asr.2005.05.048},
       adsurl = {https://ui.adsabs.harvard.edu/abs/2005AdSpR..35.1137M}
}

@ARTICLE{2013ApJ...770..103H,
       author = {{Harrison}, Fiona A. and {Craig}, William W. and {Christensen}, Finn E. and {Hailey}, Charles J. and {Zhang}, William W. and {Boggs}, Steven E. and {Stern}, Daniel and {Cook}, W. Rick and {Forster}, Karl and {Giommi}, Paolo and {Grefenstette}, Brian W. and {Kim}, Yunjin and {Kitaguchi}, Takao and {Koglin}, Jason E. and {Madsen}, Kristin K. and {Mao}, Peter H. and {Miyasaka}, Hiromasa and {Mori}, Kaya and {Perri}, Matteo and {Pivovaroff}, Michael J. and {Puccetti}, Simonetta and {Rana}, Vikram R. and {Westergaard}, Niels J. and {Willis}, Jason and {Zoglauer}, Andreas and {An}, Hongjun and {Bachetti}, Matteo and {Barri{\`e}re}, Nicolas M. and {Bellm}, Eric C. and {Bhalerao}, Varun and {Brejnholt}, Nicolai F. and {Fuerst}, Felix and {Liebe}, Carl C. and {Markwardt}, Craig B. and {Nynka}, Melania and {Vogel}, Julia K. and {Walton}, Dominic J. and {Wik}, Daniel R. and {Alexander}, David M. and {Cominsky}, Lynn R. and {Hornschemeier}, Ann E. and {Hornstrup}, Allan and {Kaspi}, Victoria M. and {Madejski}, Greg M. and {Matt}, Giorgio and {Molendi}, Silvano and {Smith}, David M. and {Tomsick}, John A. and {Ajello}, Marco and {Ballantyne}, David R. and {Balokovi{\'c}}, Mislav and {Barret}, Didier and {Bauer}, Franz E. and {Blandford}, Roger D. and {Brandt}, W. Niel and {Brenneman}, Laura W. and {Chiang}, James and {Chakrabarty}, Deepto and {Chenevez}, Jerome and {Comastri}, Andrea and {Dufour}, Francois and {Elvis}, Martin and {Fabian}, Andrew C. and {Farrah}, Duncan and {Fryer}, Chris L. and {Gotthelf}, Eric V. and {Grindlay}, Jonathan E. and {Helfand}, David J. and {Krivonos}, Roman and {Meier}, David L. and {Miller}, Jon M. and {Natalucci}, Lorenzo and {Ogle}, Patrick and {Ofek}, Eran O. and {Ptak}, Andrew and {Reynolds}, Stephen P. and {Rigby}, Jane R. and {Tagliaferri}, Gianpiero and {Thorsett}, Stephen E. and {Treister}, Ezequiel and {Urry}, C. Megan},
        title = "{The Nuclear Spectroscopic Telescope Array (NuSTAR) High-energy X-Ray Mission}",
      journal = {\apj},
         year = 2013,
        month = jun,
       volume = {770},
       number = {2},
          eid = {103},
        pages = {103},
          doi = {10.1088/0004-637X/770/2/103},
archivePrefix = {arXiv},
       eprint = {1301.7307},
 primaryClass = {astro-ph.IM}
}

@INPROCEEDINGS{1996ASPC..101...17A,
       author = {{Arnaud}, K.~A.},
        title = "{XSPEC: The First Ten Years}",
    booktitle = {Astronomical Data Analysis Software and Systems V},
         year = 1996,
       editor = {{Jacoby}, George H. and {Barnes}, Jeannette},
       series = {Astronomical Society of the Pacific Conference Series},
       volume = {101},
        month = jan,
        pages = {17}
}

@ARTICLE{Alizai2020MNRAS.494.2509A,
       author = {{Alizai}, K. and {Chenevez}, J. and {Brandt}, S. and {Lund}, N.},
        title = "{A catalogue of intermediate-duration Type I X-ray bursts observed with the INTEGRAL satellite}",
      journal = {\mnras},
         year = 2020,
        month = may,
       volume = {494},
       number = {2},
        pages = {2509-2522},
          doi = {10.1093/mnras/staa831},
archivePrefix = {arXiv},
       eprint = {2003.09324},
 primaryClass = {astro-ph.HE},
       adsurl = {https://ui.adsabs.harvard.edu/abs/2020MNRAS.494.2509A}
}

@ARTICLE{Lin2007ApJ...667.1073L,
       author = {{Lin}, Dacheng and {Remillard}, Ronald A. and {Homan}, Jeroen},
        title = "{Evaluating Spectral Models and the X-Ray States of Neutron Star X-Ray Transients}",
      journal = {\apj},
         year = 2007,
        month = oct,
       volume = {667},
       number = {2},
        pages = {1073-1086},
          doi = {10.1086/521181},
archivePrefix = {arXiv},
       eprint = {astro-ph/0702089},
 primaryClass = {astro-ph},
       adsurl = {https://ui.adsabs.harvard.edu/abs/2007ApJ...667.1073L}
}

@ARTICLE{Stoop2021MNRAS.507..330S,
       author = {{Stoop}, M. and {van den Eijnden}, J. and {Degenaar}, N. and {Bahramian}, A. and {Swihart}, S.~J. and {Strader}, J. and {Jim{\'e}nez-Ibarra}, F. and {Mu{\~n}oz-Darias}, T. and {Armas Padilla}, M. and {Shaw}, A.~W. and {Maccarone}, T.~J. and {Wijnands}, R. and {Russell}, T.~D. and {Hern{\'a}ndez Santisteban}, J.~V. and {Miller-Jones}, J.~C.~A. and {Russell}, D.~M. and {Maitra}, D. and {Heinke}, C.~O. and {Sivakoff}, G.~R. and {Lewis}, F. and {Bramich}, D.~M.},
        title = "{Multiwavelength observations reveal a faint candidate black hole X-ray binary in IGR J17285-2922}",
      journal = {\mnras},
         year = 2021,
        month = oct,
       volume = {507},
       number = {1},
        pages = {330-349},
          doi = {10.1093/mnras/stab2127},
       adsurl = {https://ui.adsabs.harvard.edu/abs/2021MNRAS.507..330S}
}

@ARTICLE{Burke2017MNRAS.466..194B,
       author = {{Burke}, M.~J. and {Gilfanov}, M. and {Sunyaev}, R.},
        title = "{A dichotomy between the hard state spectral properties of black hole and neutron star X-ray binaries}",
      journal = {\mnras},
         year = 2017,
        month = apr,
       volume = {466},
       number = {1},
        pages = {194-212},
          doi = {10.1093/mnras/stw2514},
archivePrefix = {arXiv},
       eprint = {1609.09511},
 primaryClass = {astro-ph.HE},
       adsurl = {https://ui.adsabs.harvard.edu/abs/2017MNRAS.466..194B}
}

@ARTICLE{Chen2021MNRAS.503.3540C,
       author = {{Chen}, Hai-Liang and {Tauris}, Thomas M. and {Han}, Zhanwen and {Chen}, Xuefei},
        title = "{Formation of millisecond pulsars with helium white dwarfs, ultra-compact X-ray binaries, and gravitational wave sources}",
      journal = {\mnras},
         year = 2021,
        month = may,
       volume = {503},
       number = {3},
        pages = {3540-3551},
          doi = {10.1093/mnras/stab670},
archivePrefix = {arXiv},
       eprint = {2103.02931},
 primaryClass = {astro-ph.SR},
       adsurl = {https://ui.adsabs.harvard.edu/abs/2021MNRAS.503.3540C}
}

@ARTICLE{Paczynski1981ApJ...248L..27P,
       author = {{Paczynski}, B. and {Sienkiewicz}, R.},
        title = "{Gravitational radiation and the evolution of cataclysmic binaries}",
      journal = {\apjl},
         year = 1981,
        month = aug,
       volume = {248},
        pages = {L27-L30},
          doi = {10.1086/183616},
       adsurl = {https://ui.adsabs.harvard.edu/abs/1981ApJ...248L..27P}
}

@INCOLLECTION{Bahramian2023hxga.book..120B,
       author = {{Bahramian}, Arash and {Degenaar}, Nathalie},
        title = "{Low-Mass X-ray Binaries}",
    booktitle = {Handbook of X-ray and Gamma-ray Astrophysics. Edited by Cosimo Bambi and Andrea Santangelo},
         year = 2023,
          eid = {120},
        pages = {120},
          doi = {10.1007/978-981-16-4544-0_94-1},
       adsurl = {https://ui.adsabs.harvard.edu/abs/2023hxga.book..120B}
}

@ARTICLE{Zhu2012RAA....12.1526Z,
       author = {{Zhu}, Chun-Hua and {L{\"u}}, Guo-Liang and {Wang}, Zhao-Jun},
        title = "{Population synthesis of ultra-compact X-ray binaries}",
      journal = {Research in Astronomy and Astrophysics},
         year = 2012,
        month = nov,
       volume = {12},
       number = {11},
        pages = {1526-1534},
          doi = {10.1088/1674-4527/12/11/007},
       adsurl = {https://ui.adsabs.harvard.edu/abs/2012RAA....12.1526Z}
}

@ARTICLE{Rappaport1982ApJ...254..616R,
       author = {{Rappaport}, S. and {Joss}, P.~C. and {Webbink}, R.~F.},
        title = "{The evolution of highly compact binary stellar systems.}",
      journal = {\apj},
         year = 1982,
        month = mar,
       volume = {254},
        pages = {616-640},
          doi = {10.1086/159772},
       adsurl = {https://ui.adsabs.harvard.edu/abs/1982ApJ...254..616R}
}

@INPROCEEDINGS{Verbunt1995xrbi.nasa..457V,
       author = {{Verbunt}, F. and {van den Heuvel}, E.~P.~J.},
        title = "{Formation and evolution of neutron stars and black holes in binaries.}",
    booktitle = {X-ray Binaries},
         year = 1995,
        month = jan,
        pages = {457-494},
       adsurl = {https://ui.adsabs.harvard.edu/abs/1995xrbi.nasa..457V}
}

@ARTICLE{Amaro-Seoane2023LRR....26....2A,
       author = {{Amaro-Seoane}, Pau and {Andrews}, Jeff and {Arca Sedda}, Manuel and {Askar}, Abbas and {Baghi}, Quentin and {Balasov}, Razvan and {Bartos}, Imre and {Bavera}, Simone S. and {Bellovary}, Jillian and {Berry}, Christopher P.~L. and {Berti}, Emanuele and {Bianchi}, Stefano and {Blecha}, Laura and {Blondin}, St{\'e}phane and {Bogdanovi{\'c}}, Tamara and {Boissier}, Samuel and {Bonetti}, Matteo and {Bonoli}, Silvia and {Bortolas}, Elisa and {Breivik}, Katelyn and {Capelo}, Pedro R. and {Caramete}, Laurentiu and {Cattorini}, Federico and {Charisi}, Maria and {Chaty}, Sylvain and {Chen}, Xian and {Chru{\'s}li{\'n}ska}, Martyna and {Chua}, Alvin J.~K. and {Church}, Ross and {Colpi}, Monica and {D'Orazio}, Daniel and {Danielski}, Camilla and {Davies}, Melvyn B. and {Dayal}, Pratika and {De Rosa}, Alessandra and {Derdzinski}, Andrea and {Destounis}, Kyriakos and {Dotti}, Massimo and {Du{\c{t}}an}, Ioana and {Dvorkin}, Irina and {Fabj}, Gaia and {Foglizzo}, Thierry and {Ford}, Saavik and {Fouvry}, Jean-Baptiste and {Franchini}, Alessia and {Fragos}, Tassos and {Fryer}, Chris and {Gaspari}, Massimo and {Gerosa}, Davide and {Graziani}, Luca and {Groot}, Paul and {Habouzit}, Melanie and {Haggard}, Daryl and {Haiman}, Zoltan and {Han}, Wen-Biao and {Istrate}, Alina and {Johansson}, Peter H. and {Khan}, Fazeel Mahmood and {Kimpson}, Tomas and {Kokkotas}, Kostas and {Kong}, Albert and {Korol}, Valeriya and {Kremer}, Kyle and {Kupfer}, Thomas and {Lamberts}, Astrid and {Larson}, Shane and {Lau}, Mike and {Liu}, Dongliang and {Lloyd-Ronning}, Nicole and {Lodato}, Giuseppe and {Lupi}, Alessandro and {Ma}, Chung-Pei and {Maccarone}, Tomas and {Mandel}, Ilya and {Mangiagli}, Alberto and {Mapelli}, Michela and {Mathis}, St{\'e}phane and {Mayer}, Lucio and {McGee}, Sean and {McKernan}, Berry and {Miller}, M. Coleman and {Mota}, David F. and {Mumpower}, Matthew and {Nasim}, Syeda S. and {Nelemans}, Gijs and {Noble}, Scott and {Pacucci}, Fabio and {Panessa}, Francesca and {Paschalidis}, Vasileios and {Pfister}, Hugo and {Porquet}, Delphine and {Quenby}, John and {Ricarte}, Angelo and {R{\"o}pke}, Friedrich K. and {Regan}, John and {Rosswog}, Stephan and {Ruiter}, Ashley and {Ruiz}, Milton and {Runnoe}, Jessie and {Schneider}, Raffaella and {Schnittman}, Jeremy and {Secunda}, Amy and {Sesana}, Alberto and {Seto}, Naoki and {Shao}, Lijing and {Shapiro}, Stuart and {Sopuerta}, Carlos and {Stone}, Nicholas C. and {Suvorov}, Arthur and {Tamanini}, Nicola and {Tamfal}, Tomas and {Tauris}, Thomas and {Temmink}, Karel and {Tomsick}, John and {Toonen}, Silvia and {Torres-Orjuela}, Alejandro and {Toscani}, Martina and {Tsokaros}, Antonios and {Unal}, Caner and {V{\'a}zquez-Aceves}, Ver{\'o}nica and {Valiante}, Rosa and {van Putten}, Maurice and {van Roestel}, Jan and {Vignali}, Christian and {Volonteri}, Marta and {Wu}, Kinwah and {Younsi}, Ziri and {Yu}, Shenghua and {Zane}, Silvia and {Zwick}, Lorenz and {Antonini}, Fabio and {Baibhav}, Vishal and {Barausse}, Enrico and {Bonilla Rivera}, Alexander and {Branchesi}, Marica and {Branduardi-Raymont}, Graziella and {Burdge}, Kevin and {Chakraborty}, Srija and {Cuadra}, Jorge and {Dage}, Kristen and {Davis}, Benjamin and {de Mink}, Selma E. and {Decarli}, Roberto and {Doneva}, Daniela and {Escoffier}, Stephanie and {Gandhi}, Poshak and {Haardt}, Francesco and {Lousto}, Carlos O. and {Nissanke}, Samaya and {Nordhaus}, Jason and {O'Shaughnessy}, Richard and {Portegies Zwart}, Simon and {Pound}, Adam and {Schussler}, Fabian and {Sergijenko}, Olga and {Spallicci}, Alessandro and {Vernieri}, Daniele and {Vigna-G{\'o}mez}, Alejandro},
        title = "{Astrophysics with the Laser Interferometer Space Antenna}",
      journal = {Living Reviews in Relativity},
         year = 2023,
        month = dec,
       volume = {26},
       number = {1},
          eid = {2},
        pages = {2},
          doi = {10.1007/s41114-022-00041-y},
archivePrefix = {arXiv},
       eprint = {2203.06016},
 primaryClass = {gr-qc},
       adsurl = {https://ui.adsabs.harvard.edu/abs/2023LRR....26....2A}
}

@ARTICLE{Nelemans2018arXiv180701060N,
       author = {{Nelemans}, Gijs},
        title = "{Binaries as Sources of Gravitational Waves}",
      journal = {arXiv e-prints},
         year = 2018,
        month = jul,
          eid = {arXiv:1807.01060},
        pages = {arXiv:1807.01060},
          doi = {10.48550/arXiv.1807.01060},
archivePrefix = {arXiv},
       eprint = {1807.01060},
 primaryClass = {astro-ph.SR},
       adsurl = {https://ui.adsabs.harvard.edu/abs/2018arXiv180701060N}
}

@ARTICLE{Tauris2018PhRvL.121m1105T,
       author = {{Tauris}, Thomas M.},
        title = "{Disentangling Coalescing Neutron-Star-White-Dwarf Binaries for LISA}",
      journal = {\prl},
         year = 2018,
        month = sep,
       volume = {121},
       number = {13},
          eid = {131105},
        pages = {131105},
          doi = {10.1103/PhysRevLett.121.131105},
archivePrefix = {arXiv},
       eprint = {1809.03504},
 primaryClass = {astro-ph.SR},
       adsurl = {https://ui.adsabs.harvard.edu/abs/2018PhRvL.121m1105T}
}

@ARTICLE{Nelemans2010NewAR..54...87N,
       author = {{Nelemans}, G. and {Jonker}, P.~G.},
        title = "{Ultra-compact (X-ray) binaries}",
      journal = {\nar},
         year = 2010,
        month = mar,
       volume = {54},
       number = {3-6},
        pages = {87-92},
          doi = {10.1016/j.newar.2010.09.021},
archivePrefix = {arXiv},
       eprint = {astro-ph/0605722},
 primaryClass = {astro-ph},
       adsurl = {https://ui.adsabs.harvard.edu/abs/2010NewAR..54...87N}
}

@ARTICLE{ArmasPadilla2018MNRAS.473.3789A,
       author = {{Armas Padilla}, M. and {Ponti}, G. and {De Marco}, B. and {Mu{\~n}oz-Darias}, T. and {Haberl}, F.},
        title = "{The very faint hard state of the persistent neutron star X-ray binary SLX 1737-282 near the Galactic Centre}",
      journal = {\mnras},
         year = 2018,
        month = jan,
       volume = {473},
       number = {3},
        pages = {3789-3795},
          doi = {10.1093/mnras/stx2538},
archivePrefix = {arXiv},
       eprint = {1706.01479},
 primaryClass = {astro-ph.HE},
       adsurl = {https://ui.adsabs.harvard.edu/abs/2018MNRAS.473.3789A}
}

@ARTICLE{Kubota1998PASJ...50..667K,
       author = {{Kubota}, Aya and {Tanaka}, Yasuo and {Makishima}, Kazuo and {Ueda}, Yoshihiro and {Dotani}, Tadayasu and {Inoue}, Hajime and {Yamaoka}, Kazutaka},
        title = "{Evidence for a Black Hole in the X-Ray Transient GRS 1009-45}",
      journal = {\pasj},
         year = 1998,
        month = dec,
       volume = {50},
        pages = {667-673},
          doi = {10.1093/pasj/50.6.667},
       adsurl = {https://ui.adsabs.harvard.edu/abs/1998PASJ...50..667K}
}

@ARTICLE{ArmasPadilla2017MNRAS.467..290A,
       author = {{Armas Padilla}, M. and {Ueda}, Y. and {Hori}, T. and {Shidatsu}, M. and {Mu{\~n}oz-Darias}, T.},
        title = "{Suzaku spectroscopy of the neutron star transient 4U 1608-52 during its outburst decay.}",
      journal = {\mnras},
         year = 2017,
        month = may,
       volume = {467},
       number = {1},
        pages = {290-297},
          doi = {10.1093/mnras/stx020},
archivePrefix = {arXiv},
       eprint = {1701.02728},
 primaryClass = {astro-ph.HE},
       adsurl = {https://ui.adsabs.harvard.edu/abs/2017MNRAS.467..290A}
}

@ARTICLE{armaspadilla3030_2013MNRAS.436L..89A,
       author = {{Armas Padilla}, M. and {Wijnands}, R. and {Degenaar}, N.},
        title = "{XMM-Newton and Swift spectroscopy of the newly discovered very faint  X-ray transient IGR J17494-3030.}",
      journal = {\mnras},
         year = 2013,
        month = nov,
       volume = {436},
        pages = {L89-L93},
          doi = {10.1093/mnrasl/slt119},
archivePrefix = {arXiv},
       eprint = {1307.6009},
 primaryClass = {astro-ph.HE},
       adsurl = {https://ui.adsabs.harvard.edu/abs/2013MNRAS.436L..89A}
}

@ARTICLE{zamperi1995ApJ...439..849Z,
       author = {{Zampieri}, Luca and {Turolla}, Roberto and {Zane}, Silvia and {Treves}, Aldo},
        title = "{X-Ray Spectra from Neutron Stars Accreting at Low Rates}",
      journal = {\apj},
         year = 1995,
        month = feb,
       volume = {439},
        pages = {849},
          doi = {10.1086/175223},
archivePrefix = {arXiv},
       eprint = {astro-ph/9407067},
 primaryClass = {astro-ph},
       adsurl = {https://ui.adsabs.harvard.edu/abs/1995ApJ...439..849Z}
}

@ARTICLE{Arnason2015ApJ...807...52A,
       author = {{Arnason}, R.~M. and {Sivakoff}, G.~R. and {Heinke}, C.~O. and {Cohn}, H.~N. and {Lugger}, P.~M.},
        title = "{A Low-mass Main-sequence Star and Accretion Disk in the Very Faint X-Ray Transient M15 X-3}",
      journal = {\apj},
         year = 2015,
        month = jul,
       volume = {807},
       number = {1},
          eid = {52},
        pages = {52},
          doi = {10.1088/0004-637X/807/1/52},
archivePrefix = {arXiv},
       eprint = {1505.07117},
 primaryClass = {astro-ph.HE},
       adsurl = {https://ui.adsabs.harvard.edu/abs/2015ApJ...807...52A}
}

@ARTICLE{Allen2015ApJ...801...10A,
       author = {{Allen}, Jessamyn L. and {Linares}, Manuel and {Homan}, Jeroen and {Chakrabarty}, Deepto},
        title = "{Spectral Softening Between Outburst and Quiescence In The Neutron Star Low-Mass X-Ray Binary SAX J1750.8-2900}",
      journal = {\apj},
         year = 2015,
        month = mar,
       volume = {801},
       number = {1},
          eid = {10},
        pages = {10},
          doi = {10.1088/0004-637X/801/1/10},
archivePrefix = {arXiv},
       eprint = {1501.02238},
 primaryClass = {astro-ph.HE},
       adsurl = {https://ui.adsabs.harvard.edu/abs/2015ApJ...801...10A}
}

@ARTICLE{ArmasPadilla2013MNRAS.434.1586A,
       author = {{Armas Padilla}, M. and {Degenaar}, N. and {Wijnands}, R.},
        title = "{The X-ray spectral properties of very-faint persistent neutron star X-ray binaries}",
      journal = {\mnras},
         year = 2013,
        month = sep,
       volume = {434},
       number = {2},
        pages = {1586-1592},
          doi = {10.1093/mnras/stt1114},
archivePrefix = {arXiv},
       eprint = {1303.6640},
 primaryClass = {astro-ph.HE},
       adsurl = {https://ui.adsabs.harvard.edu/abs/2013MNRAS.434.1586A}
}

@ARTICLE{saavedragx13,
       author = {{Saavedra}, Enzo A. and {Garc{\'\i}a}, Federico and {Fogantini}, Federico A. and {M{\'e}ndez}, Mariano and {Combi}, Jorge A. and {Luque-Escamilla}, Pedro L. and {Mart{\'\i}}, Josep},
        title = "{Relativistic X-ray reflection and photoionized absorption in the neutron star low-mass X-ray binary GX 13+1}",
      journal = {\mnras},
         year = 2023,
        month = jul,
       volume = {522},
       number = {3},
        pages = {3367-3377},
          doi = {10.1093/mnras/stad1157},
archivePrefix = {arXiv},
       eprint = {2304.03130},
 primaryClass = {astro-ph.HE},
       adsurl = {https://ui.adsabs.harvard.edu/abs/2023MNRAS.522.3367S}
}

@ARTICLE{ArmasPadilla2023A&A...677A.186A,
       author = {{Armas Padilla}, M. and {Corral-Santana}, J.~M. and {Borghese}, A. and {C{\'u}neo}, V.~A. and {Mu{\~n}oz-Darias}, T. and {Casares}, J. and {Torres}, M.~A.~P.},
        title = "{UltraCompCAT: A comprehensive catalogue of ultra-compact and short orbital period X-ray binaries}",
      journal = {\aap},
         year = 2023,
        month = sep,
       volume = {677},
          eid = {A186},
        pages = {A186},
          doi = {10.1051/0004-6361/202346797},
archivePrefix = {arXiv},
       eprint = {2305.07691},
 primaryClass = {astro-ph.HE},
       adsurl = {https://ui.adsabs.harvard.edu/abs/2023A&A...677A.186A}
}

@ARTICLE{Lasota2008A&A...486..523L,
       author = {{Lasota}, J. -P. and {Dubus}, G. and {Kruk}, K.},
        title = "{Stability of helium accretion discs in ultracompact binaries}",
      journal = {\aap},
         year = 2008,
        month = aug,
       volume = {486},
       number = {2},
        pages = {523-528},
          doi = {10.1051/0004-6361:200809658},
archivePrefix = {arXiv},
       eprint = {0802.3848},
 primaryClass = {astro-ph},
       adsurl = {https://ui.adsabs.harvard.edu/abs/2008A&A...486..523L}
}

@ARTICLE{Menou2002ApJ...564L..81M,
       author = {{Menou}, Kristen and {Perna}, Rosalba and {Hernquist}, Lars},
        title = "{Hydrogen-poor Disks in Compact X-Ray Binaries}",
      journal = {\apjl},
         year = 2002,
        month = jan,
       volume = {564},
       number = {2},
        pages = {L81-L84},
          doi = {10.1086/338909},
archivePrefix = {arXiv},
       eprint = {astro-ph/0111565},
 primaryClass = {astro-ph},
       adsurl = {https://ui.adsabs.harvard.edu/abs/2002ApJ...564L..81M}
}

@ARTICLE{Degennar2013ApJ...767L..31D,
       author = {{Degenaar}, N. and {Wijnands}, R. and {Miller}, J.~M.},
        title = "{A Direct Measurement of the Heat Release in the Outer Crust of the Transiently Accreting Neutron Star XTE J1709-267}",
      journal = {\apjl},
         year = 2013,
        month = apr,
       volume = {767},
       number = {2},
          eid = {L31},
        pages = {L31},
          doi = {10.1088/2041-8205/767/2/L31},
archivePrefix = {arXiv},
       eprint = {1212.1453},
 primaryClass = {astro-ph.HE},
       adsurl = {https://ui.adsabs.harvard.edu/abs/2013ApJ...767L..31D}
}

@ARTICLE{Lasota2001NewAR..45..449L,
       author = {{Lasota}, Jean-Pierre},
        title = "{The disc instability model of dwarf novae and low-mass X-ray binary transients}",
      journal = {\nar},
         year = 2001,
        month = jun,
       volume = {45},
       number = {7},
        pages = {449-508},
          doi = {10.1016/S1387-6473(01)00112-9},
archivePrefix = {arXiv},
       eprint = {astro-ph/0102072},
 primaryClass = {astro-ph},
       adsurl = {https://ui.adsabs.harvard.edu/abs/2001NewAR..45..449L}
}

@ARTICLE{Kawai1988ApJ...330..130K,
       author = {{Kawai}, N. and {Fenimore}, E.~E. and {Middleditch}, J. and {Cruddace}, R.~G. and {Fritz}, G.~G. and {Snyder}, W.~A. and {Ulmer}, M.~P.},
        title = "{X-Ray Observations of the Galactic Center by SPARTAN 1}",
      journal = {\apj},
         year = 1988,
        month = jul,
       volume = {330},
        pages = {130},
          doi = {10.1086/166461},
       adsurl = {https://ui.adsabs.harvard.edu/abs/1988ApJ...330..130K}
}

@ARTICLE{Skinner1987Natur.330..544S,
       author = {{Skinner}, G.~K. and {Willmore}, A.~P. and {Eyles}, C.~J. and {Bertram}, D. and {Church}, M.~J.},
        title = "{Hard X-ray images of the galactic centre}",
      journal = {\nat},
         year = 1987,
        month = dec,
       volume = {330},
       number = {6148},
        pages = {544-547},
          doi = {10.1038/330544a0},
       adsurl = {https://ui.adsabs.harvard.edu/abs/1987Natur.330..544S}
}

@ARTICLE{Saavedra2023A&A...680A..88S,
       author = {{Saavedra}, Enzo A. and {Fogantini}, Federico A. and {Escobar}, Gast{\'o}n J. and {Romero}, Gustavo E. and {Combi}, Jorge A. and {Marcel}, Estefania},
        title = "{NuSTAR and XMM-Newton observations of the binary 4FGL J1405.1-6119. A {\ensuremath{\gamma}}-ray-emitting microquasar?}",
      journal = {\aap},
         year = 2023,
        month = dec,
       volume = {680},
          eid = {A88},
        pages = {A88},
          doi = {10.1051/0004-6361/202347760},
archivePrefix = {arXiv},
       eprint = {2310.11553},
 primaryClass = {astro-ph.HE},
       adsurl = {https://ui.adsabs.harvard.edu/abs/2023A&A...680A..88S}
}

@ARTICLE{VanHaaften2013A&A...552A..69V,
       author = {{van Haaften}, L.~M. and {Nelemans}, G. and {Voss}, R. and {Toonen}, S. and {Portegies Zwart}, S.~F. and {Yungelson}, L.~R. and {van der Sluys}, M.~V.},
        title = "{Population synthesis of ultracompact X-ray binaries in the Galactic bulge}",
      journal = {\aap},
         year = 2013,
        month = apr,
       volume = {552},
          eid = {A69},
        pages = {A69},
          doi = {10.1051/0004-6361/201220552},
archivePrefix = {arXiv},
       eprint = {1302.7181},
 primaryClass = {astro-ph.SR},
       adsurl = {https://ui.adsabs.harvard.edu/abs/2013A&A...552A..69V}
}

@ARTICLE{Wijnands2015MNRAS.454.1371W,
       author = {{Wijnands}, R. and {Degenaar}, N. and {Armas Padilla}, M. and {Altamirano}, D. and {Cavecchi}, Y. and {Linares}, M. and {Bahramian}, A. and {Heinke}, C.~O.},
        title = "{Low-level accretion in neutron star X-ray binaries}",
      journal = {\mnras},
         year = 2015,
        month = dec,
       volume = {454},
       number = {2},
        pages = {1371-1386},
          doi = {10.1093/mnras/stv1974},
archivePrefix = {arXiv},
       eprint = {1409.6265},
 primaryClass = {astro-ph.HE},
       adsurl = {https://ui.adsabs.harvard.edu/abs/2015MNRAS.454.1371W}
}

\end{document}